\documentclass[12pt]{article}
\pdfoutput=1 

\usepackage{cancel,slashed}
\usepackage{bbm}
\usepackage{amssymb}
\usepackage{amsfonts}
\usepackage{amsmath}
\usepackage{graphicx}
\usepackage{cite}

\usepackage{latexsym}
\usepackage{color}
\usepackage{xypic}
\usepackage{cancel,slashed}
\usepackage[linktocpage]{hyperref}
\usepackage{color}
\usepackage{transparent}
\usepackage{footmisc}

\makeatletter
\def\@xfootnote[#1]{%
  \protected@xdef\@thefnmark{#1}%
  \@footnotemark\@footnotetext}
\makeatother

\newcommand{\bmat}{\left(\begin{array}}
\newcommand{\emat}{\end{array}\right)}

\def\p{\partial}

\def\g{\gamma}
\def\d{\delta}
\def\th{\theta}

\def\vphi{\varphi}
\def\-{\hphantom{-}}

\def\s2{\frac{1}{\sqrt2}}

\def\oh{\frac{1}{2}}
\def\beq{\begin{equation}}
\def\eeq{\end{equation}}
\def\beqa{\begin{eqnarray}}
\def\eeqa{\end{eqnarray}}

\def\tr{{\rm tr \,}}

\def\Dsl{\,\raise.15ex\hbox{/}\mkern-13.5mu D} 

\def\CR {{\cal R}}

\def\CO {{\cal O}}

\def\tr{\mbox{Tr}}
\def\str{\mbox{STr}}

\def\be{\begin{equation}}
\def\ee{\end{equation}}
\def\bea{\begin{eqnarray}}
\def\eea{\end{eqnarray}}
\def\raw{\rightarrow}

\def\IC{\mathbb{C}}

\def\oh{\frac{1}{2}}

\def\d{{\delta}}
\def\eps{{\epsilon}}
\def\th{{\theta}}

\def\lam{{\lambda}}

\def\g{{\gamma}}

\def\vphi{{\varphi}}
\def\p{{\partial}}

\def\vec#1{{\overrightarrow{#1}}}
\def\str{\mbox{STr}}

\def\w{{\wedge}}

\def\sm2{{\mbox{\small 2}}}

\begin{document}
\pagestyle{plain}

\makeatletter
\@addtoreset{equation}{section}
\makeatother
\renewcommand{\theequation}{\thesection.\arabic{equation}}
\pagestyle{empty}
\rightline{ IFT-UAM/CSIC-15-133}
\vspace{0.5cm}
\begin{center}
\LARGE{{Fitting fermion masses and mixings\\ 
in F-theory GUTs}
\\[10mm]}
\large{Federico Carta,\footnote[$\dagger$]{La Caixa-Severo Ochoa Scholar}$^1$ Fernando Marchesano$^1$ and Gianluca Zoccarato$^{1,2}$ \\[10mm]}
\small{
${}^1$ Instituto de F\'{\i}sica Te\'orica UAM-CSIC, Cantoblanco, 28049 Madrid, Spain \\[2mm] 
${}^2$ Departamento de F\'{\i}sica Te\'orica, 
Universidad Aut\'onoma de Madrid, 
28049 Madrid, Spain
\\[8mm]} 
\small{\bf Abstract} \\[5mm]
\end{center}
\begin{center}
\begin{minipage}[h]{15.0cm} 

We analyse the structure of Yukawa couplings in local SU(5) F-theory models with $E_7$ enhancement. These models are the minimal setting in which the whole flavour structure for the MSSM charged fermions is encoded in a small region of the entire compactification space. In this setup the $E_7$ symmetry is broken down to SU(5) by means of a 7-brane T-brane background, and further to the MSSM gauge group by means of a hypercharge flux that also implements doublet-triplet splitting. At tree-level only one family of quarks and charged leptons is massive, while the other two obtain hierarchically smaller masses when stringy non-perturbative effects are taken into account. We find that there is a unique $E_7$ model with such hierarchical flavour structure. The relative simplicity of the model allows to perform the computation of Yukawa couplings for a region of its parameter space wider than previous attempts, obtaining realistic fermion masses and mixings for large parameter regions. Our results are also valid for local models with $E_8$ enhancement, pointing towards a universal structure to describe realistic fermion masses within this framework.

\end{minipage}
\end{center}
\newpage
\setcounter{page}{1}
\pagestyle{plain}
\renewcommand{\thefootnote}{\arabic{footnote}}
\setcounter{footnote}{0}


\tableofcontents


\section{Introduction}
\label{s:intro}

A key pledge of F-theory GUT models \cite{bhv1,bhv2,dw1,dw2} is to provide a UV complete framework where gauge coupling unification and realistic Yukawa couplings occur at the same time. This a clear advantage with respect to type II string compactifications, but perhaps not so much with respect to heterotic constructions where these two features are already built in. What F-theory does provide with respect to heterotic compactifications is the advantage of computability for these Yukawa couplings. Indeed, in the context of F-theory GUTs computing physical Yukawa couplings does not require precise information over the whole compactification manifold, but rather over a submanifold $S_{\rm GUT}$ where the GUT degrees of freedom are localised. In fact, one may oftentimes only require information around a local patch $U_p \subset S_{\rm GUT}$ around a point $p$ where a certain Yukawa coupling is generated. One may even conceive a configuration where all the fermion masses and mixings angles of the MSSM arise from the same patch $U \subset S_{\rm GUT}$ \cite{Heckman:2009mn}, which is a huge simplification with respect to the heterotic case.

Such ultra-local approach is based on the description of the GUT degrees of freedom by means 8d super Yang-Mills theory localised on $S_{\rm GUT}$, a framework that has been largely used in the computation of F-theory Yukawa couplings \cite{hv08,Hayashi:2009ge,Randall:2009dw,Font:2009gq,cchv09,Conlon:2009qq,hktw2,mm09,Leontaris:2010zd,cchv10,Chiou:2011js,afim11,fimr12,fmrz13,mrz15}. In this description each MSSM particle has an internal wavefunction whose profile depends on the parameters that describe the local model. Since the Yukawa couplings are computed through the triple overlap of such wavefunctions, they will also depend on these local model parameters. Hence, given a local model, one may wonder for which parameter values one may obtain a realistic set of fermion masses and mixings. 

A more ambitious but natural question is how generic are realistic Yukawa couplings in the context of the string landscape. In this sense F-theory GUT constructions also provide a unique framework to formulate this problem. Indeed, just like in the bottom-up approach of \cite{Aldazabal:2000sa} one may embed a local F-theory GUT model with realistic Yukawas into one or several compact Calabi-Yau four-folds. The usual techniques of moduli stabilisation for F-theory vacua \cite{Grana:2005jc,Douglas:2006es,Denef:2007pq,Denef:2008wq,Quevedo:2014xia} will select a discrete set of values for the complex and K\"ahler moduli of such four-folds, which will be perceived as a discrete set of values for the parameters that describe the local model. One may then compare how many of these points correspond to realistic Yukawas in the 4d effective theory and how many do not. 

Nevertheless, addressing this problem requires of a local model where realistic Yukawas are obtained for a reasonably wide region of its parameter space, a crucial ingredient which is so far missing. Indeed, previous attempts like \cite{fimr12,fmrz13,mrz15} succeeded in providing (ultra)local models whose Yukawa couplings for charged fermions had an appealing hierarchical structure, which was in turn translated into a realistic fermion mass spectrum for certain values of the local model parameters. There was however never an accurate notion on how large is the region of parameter space that satisfies all the necessary constraints for such realistic spectrum, this partly due to the complicated formulas that describe physical Yukawas. 

The purpose of this paper is to fill this gap and provide an (ultra)local model in which realistic Yukawas are achieved for a reasonable region of its parameter space. Despite the large amount of parameters that a local model may have this feature is not that easy to obtain, as there are many constraints that this model must satisfy in order to have a suitable local chiral spectrum and realistic Yukawa couplings for it. Such constraints are oftentimes not compatible with each other, making the model not viable phenomenologically. This fact is illustrated by the recent analysis in \cite{mrz15}, which showed that out of several initial local models with similar features most of them could be discarded at the level of the holomorphic Yukawa couplings. 

To achieve this task we will follow the philosophy in \cite{Heckman:2009mn} and consider $SU(5)$ ultra-local models where the Yukawas for up and down-type quarks arise from a single patch $U \subset S_{\rm GUT}$, an assumption motivated experimentally by the CKM matrix. The simplest models with this attribute are those with an underlying $E_7$ symmetry around the patch $U$, so the first step will be to consider $E_7$ models with a realistic pattern for their Yukawas. For this we will follow the scheme in \cite{mm09} (see also \cite{afim11,fimr12,fmrz13,mrz15}) in which rank one Yukawa matrices are obtained at tree-level and then enhanced to rank three when external non-perturbative effects (like Euclidean D3-brane instantons) are taken into account.\footnote{For other approaches to the generation of Yukawa hierarchies in F-theory GUTs see e.g.  \cite{fi08,DudasPalti,Leontaris:2010zd,Krippendorf,Krippendorf:2010hj,Bizet:2014uua,Krippendorf:2015kta}.} 

In this context we will find that there is a unique $E_7$ model with a realistic pattern of Yukawa couplings, which we then analyse in detail. Interestingly, we obtain that such model contains the same structure of matter curves and Yukawa couplings as the $E_8$ model selected in \cite{mrz15} through the same criteria of realistic fermion mass spectrum. In particular we find that the Yukawa couplings for the fermions charged under the Standard Model gauge group have the same parametric dependence as those of the $E_8$ model in \cite{mrz15}. Therefore our present result can also be directly applied to local models with $E_8$ enhancement. More surprisingly, this point towards a universal structure of matter curves and Yukawa couplings in order to successfully implement the proposal in \cite{mm09} to generate flavour hierarchies, something that would have been initially hard to guess. 

In our analysis we compute the physical Yukawa couplings for this model for a wider range of parameters than compared to previous attempts, scanning for regions that are compatible with an MSSM extension of the Standard Model and their corresponding Yukawas at the GUT scale. The parameter space that we explore is not infinite, because certain quantities like local flux densities are constrained in order to obtain a realistic local chiral spectrum within the patch $U$ \cite{Palti:2012aa}. In particular we locally impose the doublet-triplet splitting mechanism of \cite{bhv2},\footnote{This  mechanism has been analysed in \cite{Dudas:2010zb,Marsano:2009wr,Marsano:2009gv,Marsano:2010sq} in the context of GUT models with a spectral cover, finding some tension with the presence of chiral exotics. Such results do not in principle apply to the models analysed here, which feature the presence of T-branes \cite{hktw2,cchv10,Donagi:2011jy,Anderson:2013rka,Collinucci:2014qfa} in their construction.} which take us to a region of the parameter space previously unexplored. Quite remarkably, it is in this region where we find that fitting fermion masses and mixing angles is easier than the analysis made in \cite{afim11,fimr12,fmrz13,mrz15}, obtaining a reasonably wide region of parameter space that corresponds to realistic Yukawas.

The following sections are organised as follows. In section \ref{s:yukawas} we review the standard construction of (ultra)local F-theory GUTs and several key results regarding their Yukawa couplings. In section \ref{s:e7models} we discuss the different local SU(5) models with $E_7$ enhancement and select the one which is promising to obtain realistic Yukawas. We then analyse this model in detail in section \ref{s:e7model} obtaining its holomorphic Yukawa couplings. We then compute the normalisation factors in section \ref{s:norm}, which allow to obtain the physical Yukawas to be discussed in section \ref{s:fitting} in the context of realistic fermion masses and CKM matrix. We draw our conclusions in section \ref{s:conclu}, relegating several technical details to the appendices.

\section{Yukawas and exceptional groups in F-theory GUTs}
\label{s:yukawas}

The standard scheme of F-theory GUT models \cite{bhv1,bhv2,dw1,dw2} (see \cite{thebook,ftheoryreviews} for reviews) requires a Calabi-Yau fourfold elliptically fibered over a three-fold base $B$, such that the fibre degenerates over a four-cycle $S_{\rm GUT} \subset B$. At such locus the Dynkin diagram of the fibre singularity corresponds to the Lie group $G_{\rm GUT}$, except at subloci like complex curves $\Sigma \subset S_{\rm GUT}$ and their intersections where the fibre exhibits a higher singularity type. A quite powerful feature that arises out of this geometric picture is that of localisation of GUT degrees of freedom. Indeed, one finds that the 4d gauge bosons that generate the gauge group have an internal profile localised at the four-cycle $S_{\rm GUT}$, and that the curves $\Sigma$ further localise 4d chiral matter charged under $G_{\rm GUT}$. This statement remains true when one adds a four-form flux $G_4$ threading $S_{\rm GUT}$ which specifies the 4d chiral matter content of the model and, if chosen appropriately, breaks $G_{\rm GUT}$ to the subgroup $SU(3) \times SU(2) \times U(1)_Y$ and implements a double-triple splitting mechanism.

This feature of localisation is easier to detect with an alternative description of the degrees of freedom localised at $S_{\rm GUT}$, which uses a 8d action related to the 7-branes wrapping $S_{\rm GUT}$ and those intersecting them at matter curves. Such action is defined on a four-cycle $S$ and in terms of a non-Abelian symmetry group $G$ that contains $G_{\rm GUT}$ and all the enhanced symmetry groups at the matter curves and their intersections. Under this description the 4d effective theory that corresponds to the GUT sector of the compactification can be obtained upon dimensional reduction of the 8d action. In particular  the computation of the Yukawa couplings is encoded in terms of the superpotential 
\be
W\, =\, m_*^4 \int_{S} \tr \left( F \wedge \Phi \right)
\label{supo7}
\ee
and the D-term
\be
D\, =\, \int_S \omega \wedge F + \frac{1}{2}  [\Phi, \Phi^\dagger]\, .
\label{FI7}
\ee
Here $m_*$ is the F-theory characteristic scale, $F = dA - i A \wedge A$ is the field strength of the 7-branes gauge boson $A$, and $\Phi$ is the so-called Higgs field: a (2,0)-form on the four-cycle $S$ describing the 7-branes transverse geometrical deformations. Both $A$ and $\Phi$ transform in the adjoint of the initial gauge group $G$, which is nevertheless broken to a subgroup due to their non-trivial profile. On the one hand the profile $\langle \Phi \rangle$ is such that it only commutes with the generators of $G_{\rm GUT}$ in the bulk of $S_{\rm GUT}$, while on top of the matter curves of $S_{\rm GUT}$ it also commutes with further roots of $G$. On the other hand the profile $\langle A \rangle$ is such that it further breaks $G_{\rm GUT}$ to the MSSM gauge group through a component along the hypercharge generator. These profiles are not arbitrary but need to solve the equations of motion that arise from minimisation of (\ref{supo7}) and (\ref{FI7}). Similarly, given the background for $\Phi$ and $A$ one can compute the zero mode equations for their fluctuations via the two functionals $W$ and $D$, and then plug them into (\ref{supo7}) to obtain the Yukawa coupling through a triple wavefunction overlap. 

With this alternative description one can extract several key features regarding the computation of Yukawa couplings in F-theory GUTs:

\begin{itemize}

\item If we consider $G_{\rm GUT}=SU(5)$ (as we will do in the following) up-type Yukawas $\mathbf{10 \times 10 \times 5}$ will arise from (\ref{supo7}) only if $G$ contains the exceptional group $E_6$.

\item The holomorphic piece of the Yukawas does not depend on the worldvolume flux profile $\langle F \rangle$, but only on the geometry around the intersection of the corresponding matter curves \cite{cchv09}. Therefore one can compute holomorphic Yukawas by specifying $\langle \Phi \rangle$ on a neighbourhood $U_p \subset S$ of the matter curves intersection point $p$.

\item The flux $\langle F \rangle$ localises the internal zero mode wavefunctions at particular regions of the matter curves. If the MSSM fields are sufficiently peaked within a patch $U \subset S$ one can compute their physical Yukawas by knowing $\langle \Phi \rangle$ and $\langle F \rangle$ in this patch and replacing $S \raw U$ in (\ref{supo7}) and (\ref{FI7}) \cite{Palti:2012aa}. 

\item For $G= E_7$ or $E_8$ all the Yukawa couplings for charged MSSM fermions can be described from a single patch $U$, a scheme favoured by the empirical values of the CKM matrix \cite{Heckman:2009mn}. 

\item One can engineer GUT models where the Yukawa matrices are of rank one by imposing a topological condition on the matter curves \cite{cchv10}. However, this is only compatible with well-defined zero mode wavefunctions if one considers a non-Abelian background profile for $ \Phi $  \cite{hktw2,cchv10}, dubbed T-brane background.

\end{itemize}

All these results point to a very suggestive setup for an F-theory GUT model, in which all the Yukawa couplings of the MSSM charged fermions are originated from a patch $U$ where $G = E_7$ or $E_8$. The Yukawa matrices are of rank one, and therefore a mass hierarchy is generated between the third and the first two families of quarks and leptons, which at this point are massless. 

Following \cite{mm09}, one may now take into account the non-perturbative effects originated at a different four-cycle $S_{\rm np} \subset B$, that modify the 7-brane superpotential to 
\be
W\, =\, m_*^4\,  \int_S \tr \left( F \wedge \Phi \right) + \eps\, \frac{\theta_0}{2} \tr \left( F \wedge F\right)
\label{suponp}
\ee
with $\eps$ measuring the strength of the non-perturbative effect and $\theta_0$ a holomorphic function that depends on the embedding of the four-cycle $S_{\rm np}$.\footnote{Namely $\theta_0\, =\,  (4\pi^2 m_*)^{-1} [{\rm log\, } h/h_0]_{S}$, with $h$ a divisor function such that $S_{\rm np} = \{h = 0\}$, and $h_0 = \int_S h$. The full superpotential contains additional terms of the form $\theta_k \str (\Phi_{xy}^k F^2)$, $k\geq2$. These additional terms are suppressed by additional terms of $m_*$  and therefore negligible compared to the contribution coming from $\theta_0$. We refer to \cite{mm09,afim11,fimr12} for further details.} This deformed superpotential will generically increase the rank of the Yukawa matrices from one to three, as has been shown by the explicit analysis of several cases of interest \cite{afim11,fimr12,fmrz13,mrz15}. One can summarise the results of this approach as follows:
 
 \begin{itemize}
 
 \item The hierarchy of fermion masses between different families can already be seen at the level of holomorphic Yukawas and in terms of the parameter $\eps$. Typically one either obtains a hierarchy of the form $(\CO(1), \CO(\eps), \CO(\eps^2))$ or $(\CO(1), \CO(\eps^2), \CO(\eps^2))$, depending on the structure of matter curves around their intersection point. 
 
 \item From these two possibilities only the pattern $(\CO(1), \CO(\eps), \CO(\eps^2))$ allows to reproduce a realistic mass spectrum for charged fermions, with a typical value of $\eps \sim 10^{-4}$. The precise fit with empiric data depends of the worldvolume flux densities threading the curves around their intersection points, and on the mass running from the TeV to the compactification scale. The latter usually selects tan $\beta \sim 20 - 50$.
 
 \item The departure from the GUT mass relations is mainly due to the dependence of the physical Yukawas on the hypercharge flux, whose effect is different for each family. Obtaining an appreciable effect entails an hypercharge flux density which is non-negligible in units of $m_*$. 
  
 \item Fermion masses are complicated functions of the flux densities that arise from the components of $\langle F \rangle$, and which in this local approach are treated as parameters. Nevertheless, mass ratios display a much simpler dependence in only a few of these parameters.  
 
 \item If the patch $U$ contains both intersection points $p_{\rm up}$ and $p_{\rm down}$ where up and down-like Yukawas are respectively generated one can also compute the CKM matrix of quark mixing angles in this local approach. The observed mixing between the third and second families constrains $p_{\rm up}$ and $p_{\rm down}$ to be very close to each other compared to the size of $S_{\rm GUT}$, pointing to a symmetry group $G$ which is either $E_7$ or $E_8$. 
 
 \end{itemize}

This last point was analysed quantitatively in \cite{mrz15} for the case of a $SU(5)$ model with symmetry group $E_8$. Such class of local models have been highlighted in \cite{Heckman:2009mn} as a tantalising possibility to generate all fermion masses (including the neutrino sector) from a local patch of the compactification. It was then seen in \cite{mrz15} that such proposal could be made compatible with the above scheme to generate hierarchical Yukawas via non-perturbative effects, at least for the sector of fermions charged under the MSSM gauge group. However a realistic fermion mass spectrum would not happen automatically, but only for certain choices of matter curves/T-brane backgrounds. 

In the following we would like to see if this scheme for generating flavour hierarchies can also be applied to models with $E_7$ enhancement. Notice that in principle  a patch of $E_7$ enhancement does not describe how the masses for the neutrino sector of a F-theory GUT model may be generated \cite{Heckman:2009mn}. Nevertheless, from the viewpoint of the approach in \cite{afim11,fimr12,fmrz13,mrz15} which only deals with Yukawas for the MSSM charged fermions, these models are as equally compelling as the ones with $E_8$ enhancement. Moreover these models are simpler in the sense that they admit fewer matter curve embeddings than the $E_8$ case. In fact, in the next section we will see that imposing rank-one Yukawa matrices at tree-level selects a unique model of $E_7$ enhancement, which due to its simplicity will be analysed in great detail in the subsequent sections.

\section{$SU(5)$ models with $E_7$ enhancement}
\label{s:e7models}

As mentioned in the introduction, one of the interesting features of $SU(5)$ models with $E_7$ enhancement is that they are rather universal, in the sense that there are very few ways to embed $SU(5)$ into $E_7$ in an F-theory construction. In fact, we will see that in this context there is essentially only one possibility to generate a hierarchical pattern of Yukawa couplings by means of the mechanism proposed in \cite{mm09}. Recall that in this scheme one needs to consider an exceptional group $E_n$ Higgsed by a T-brane background such that the resulting pattern of matter curves can embed the full chiral content of the MSSM. The T-brane profile should also be such that only one family of quarks and leptons develops non-trivial Yukawa couplings from the tree-level superpotential \cite{cchv10}. Finally,  by including non-perturbative effects the remaining families will also develop Yukawa couplings, creating a hierarchy of masses between families. If this hierarchy is of the form $(1,\eps,\eps^2)$, with $\eps$ a small number that measures the strength of the non-perturbative effect, then one obtains Yukawa matrices at the GUT scales that are suitable to reproduce experimental masses for charged fermions \cite{afim11,fimr12,fmrz13,mrz15}. 

As in \cite{mrz15} one may classify the different embeddings of $SU(5)_{\rm GUT}$ into $E_n$ by looking at the pattern of matter curves of the local models, which is in turn specified in terms of the Higgs background $\langle \Phi \rangle$. Such background takes values in the algebra $\mathfrak{g}_\perp$ defined such that $\mathfrak{g}_{\rm GUT} \oplus \mathfrak{g}_\perp$ is a maximal subalgebra of $\mathfrak{g}_p = {\rm Lie} (G_p)$. In our case $\mathfrak{g}_p = \mathfrak{e}_7$ and $\mathfrak{g}_\perp = \mathfrak{su}_3 \oplus \mathfrak{u}_1$, so the maximal decomposition of the adjoint representation reads
\bea
\label{maximal}
\mathfrak{e}_7 & \supset & \mathfrak{su}_{5}^{\rm GUT} \oplus \mathfrak{su}_{3}  \oplus \mathfrak{u}_{1} \\ \nonumber
\textbf{133} & \rightarrow & (\textbf{24},\textbf{1})_0 \oplus (\textbf{1},\textbf{8})_0   \oplus  (\textbf{1},\textbf{1})_0 \oplus (\textbf{10},\overline{\textbf{3}})_{-1} \oplus ({\textbf{5}},\overline{\textbf{3}})_{2}\oplus ({\textbf{5}},\overline{\textbf{1}})_{-3}\oplus  c.c.
\eea
By construction $\langle \Phi \rangle$ commutes with $\mathfrak{su}_{5}^{\rm GUT}$, but it acts non-trivially on the representations $\CR$ of $\mathfrak{g}_\perp =\mathfrak{su}_{3} \otimes \mathfrak{su}_{1}$ that appear as $(\CR_{\rm GUT}, \CR)$ in  (\ref{maximal}). This action can be expressed in terms of  a matrix $\Phi_{\mathcal{R}}$ such that $[\langle \Phi\rangle, \CR] = \Phi_\CR \CR$. Then at the locus where ${\rm det\, } \Phi_\CR = 0$ there will be a matter curve hosting zero modes in the representation $\CR_{\rm GUT}$ of $\mathfrak{su}_{5}^{\rm GUT}$.

One may now classify different profiles for $\langle \Phi \rangle$ in terms of the block diagonal structure of the matrices $\Phi_{\mathcal{R}}$, which we assume reconstructible in the sense of \cite{cchv10}. Because $\mathfrak{g}_\perp$ factorises as $\mathfrak{su}_{3} \otimes \mathfrak{u}_{1}$, we may directly focus on their block diagonal structure within $\mathfrak{su}_{3}$.
In order to discuss the block diagonal structure of the Higgs field it is convenient to choose $\mathcal R = \mathbf 3$, the fundamental representation of $SU(3)$
as the action of the Higgs field on any other representation may be constructed by taking suitable tensor products of the fundamental representation.
With this choice the three different possibilities we have are
%
\begin{enumerate}
\item[{\it i)}] $\Phi_{\bf 3}$ is diagonal 
\item[{\it ii)}] $\Phi_{\bf 3}$ has a $2+1$ block structure 
\item[{\it iii)}] $\Phi_{\bf 3}$ has a single block
\end{enumerate}
Out of these three options the first one represents a $\langle \Phi \rangle$ taking values in the Cartan subalgebra of $\mathfrak{e}_7$, and so it does not correspond to a T-brane background. Option {\it iii)} was analysed in \cite{Chiou:2011js}, obtaining that up-type Yukawa couplings identically vanish. Hence, we are left with a splitting of the form  {\it ii)} as the only possibility to obtain realistic hierarchical pattern of Yukawa couplings. 

Reconstructible models with the split $2+1$ can be characterised with a profile for $\langle \Phi \rangle$ lying in the subalgebra $\mathfrak{su}_{2} \oplus \mathfrak{u}_{1} \subset \mathfrak{su}_{3} \subset \mathfrak{g}_\perp$. Hence in order to read the spectrum of matter curves one may adapt the above branching rules for the adjoint of $\mathfrak{e}_7$ to the non-maximal decomposition 
$\mathfrak{su}_{5}^{\rm GUT} \oplus \mathfrak{su}_{2}  \oplus \mathfrak{u}_{1} \oplus \mathfrak{u}_{1}$. We obtain\footnote{In writing the decomposition of the $\mathfrak{e}_7$ Lie algebra under $\mathfrak{su}_{5}^{\rm GUT} \oplus \mathfrak{su}_{2}  \oplus \mathfrak{u}_{1} \oplus \mathfrak{u}_{1}$ we choose two particular combinations of the generators of $\mathfrak{u}_1 \oplus \mathfrak{u}_1$.}
%
%
%
\bea
\label{nonmaximal}
\mathfrak{e}_7 & \supset & \mathfrak{su}_{5}^{\rm GUT} \oplus \mathfrak{su}_{2}  \oplus \mathfrak{u}_{1} \oplus \mathfrak{u}_{1}\\ \nonumber
\textbf{133} & \rightarrow & (\textbf{24},\textbf{1})_{0,0} \oplus (\textbf{1},\textbf{3})_{0,0}  \oplus  2(\textbf{1},\textbf{1})_{0,0} \oplus ( (\textbf{1},\textbf{2})_{-2,1} \oplus  c.c.)  \\ \nonumber 
& & \oplus\ (\textbf{10},{\textbf{2}})_{1,0} \oplus (\textbf{10},{\textbf{1}})_{-1,1} \oplus ({\textbf{5}},{\textbf{2}})_{0,-1} \oplus ({\textbf{5}},{\textbf{1}})_{-2,0} \oplus ({\textbf{5}},{\textbf{1}})_{1,1}\oplus  c.c.
\eea
and so we have two different kinds of {\bf 10} matter curves and three kinds of {\bf 5} matter curves. In order to have a rank one up-type Yukawa matrix we need to identify the matter curve ${\bf 10}_M$ with $ (\textbf{10},{\textbf{2}})_{1,0}$. Hence the curve containing the Higgs up is fixed to be $({\textbf{5}},{\textbf{1}})_{-2,0}$, or otherwise the Yukawa coupling ${\bf 10}_M \times {\bf 10}_M \times {\bf 5}_U$ cannot be generated. Finally, the remaining two {\bf 5}-curves must host the family representations $\bar{\bf 5}_M$ and down Higgs representation $\bar{\bf 5}_D$, respectively. 

To summarise, we find that in order to obtain a hierarchical pattern of Yukawa couplings we only have two possible ways to identify the matter curves with the representations of $SU(5)_{\rm GUT}$. Namely those are:
\begin{enumerate}
\item {\bf Model A}\\
\begin{equation}
\begin{aligned}
(\mathbf{10},\mathbf{2})_{1,0}&=\mathbf{10}_M\\
(\mathbf 5,\mathbf 1)_{-2,0}&=\mathbf{5}_U\\
(\mathbf{\bar{5}},\mathbf 2)_{0,1}&=\mathbf{\bar{5}}_M\\
(\mathbf{\bar{5}},\mathbf 1)_{-1,-1}&=\mathbf{\bar{5}}_D
\end{aligned}
\end{equation}

\item {\bf Model B}\\
\begin{equation}
\begin{aligned}
(\mathbf{10},\mathbf 2)_{1,0}&=\mathbf{10}_M\\
(\mathbf 5,\mathbf 1)_{-2,0}&=\mathbf{5}_U\\
(\mathbf{\bar{5}},\mathbf{1})_{-1,-1}&=\mathbf{\bar{5}}_M\\
(\mathbf{\bar{5}},\mathbf 2)_{0,1}&=\mathbf{\bar{5}}_D
\end{aligned}
\end{equation}
\end{enumerate}

In the following we will compute the Yukawa couplings for both of these scenarios. Even if we obtain a favourable hierarchical pattern $(1, \eps, \eps^2)$ for both of them, we advance that only Model A will reveal itself physically viable. Therefore we will discuss our results mainly in terms of this first case, deferring many computational details regarding Model B to Appendix \ref{ap:modelB}.

\section{Yukawa hierarchies in the $E_7$ model}
\label{s:e7model}

Let us now consider in more detail the two models with a local $E_7$ enhancement highlighted in the previous section. 
Since the difference between them amounts to how matter fields are distributed among matter curves, it is possible to give 
a description of the local background for the Higgs field $\Phi$ and the gauge connection $A$ that applies to both models 
at the same time.

Indeed, such local models with $E_7$ enhancement is specified by choosing a Higgs field $\Phi$ and a gauge connection $A$ valued in the algebra $\mathfrak{su}_2
\oplus \mathfrak{u}_1\oplus \mathfrak{u}_1$. To preserve supersymmetry in the low-energy 4d theory it is necessary to choose background fields satisfying
the following supersymmetry equations
\begin{subequations}
\label{Fterm7}
\begin{align}
\bar \partial_{A} \Phi = &\, 0\\
F^{(0,2)} =&\, 0\\
\omega \wedge F +\frac{1}{2} [\Phi, \Phi^\dagger] =&\,0\,.
\end{align}
\end{subequations}
The first two equations ensure the vanishing of the F-terms and may be obtained by varying the superpotential (\ref{supo7}) while the third equation ensures the vanishing
of the D-term (\ref{FI7}).\footnote{When we consider the corrected superpotential (\ref{suponp}) these F-terms equations will be modified, shifting the background values for $\Phi$ and $A$ \cite{fimr12,fmrz13}. This shift will be taken into account when computing the zero mode wavefunctions in section \ref{sec:residuemain}.} A common strategy to find a solution of the previous set of equation is to exploit the fact that the F-term equations are invariant under 
complexified gauge transformations. In particular this gives the possibility of fixing a particular gauge, usually called \emph{holomorphic gauge}, where 
the gauge connection satisfies $A^{(0,1)}=0$. In this gauge the F-term equations greatly simplify and any choice of holomorphic Higgs field is a solution. While this solution is not a physical one (in the sense that the gauge connection is not real and the D-term equations are not satisfied) it still
gives insight on the structure of matter curves and the rank of the Yukawa matrix. To reach a physically sensible solution of the equations 
of motion we may perform a complexified gauge transformation that brings the gauge fields in a real gauge that also satisfies the D-term equations. This
is a rather cumbersome task in models with T-branes but nevertheless it is a necessary step to extract the physical values of the Yukawa couplings. 

With this approach in mind we will start introducing the background value of the Higgs field in holomorphic gauge discussing moreover the structure of the
various matter curves. After this we will consider the passage to a real gauge and impose the D-term equations. This will force the introduction of some
fluxes which are non-primitive and leaves open the possibility to add primitive fluxes. We will discuss the addition of these fluxes that allow for a chiral 
spectrum in the 4d theory and the breaking of the GUT group down to the MSSM gauge group. We close this section with a direct computation of the 
Yukawa matrices for both models introduced in the previous section.

\subsection{Higgs background}
\subsubsection*{Holomorphic gauge}

The first element that enters in the definition of our local model is the vacuum expectation value of the Higgs field $\langle \Phi \rangle =
\langle \Phi_{xy} \rangle \, dx \wedge dy$ which constitutes the primary source of breaking the symmetry group $E_7$ down to $SU(5)_{GUT}$. Our choice in holomorphic
gauge is the following one
\be
\langle \Phi_{xy} \rangle = m\left(E^++mxE^-\right)+{\mu}_1^2\left({a}x-y\right)Q_1+\left[{\mu}_2^2\left({b}x-y\right)+\kappa\right]Q_2
\label{Phixy}
\ee
where $Q_i$ and $E^{\pm}$ are generators of the Lie algebra of $E_7$ whose definition (along with other details involving the $E_7$ Lie algebra) are 
given in appendix \ref{ap:e7}. In the definition of the Higgs background we introduced the complex constants $m$, $\mu_{1,2}$ and $\kappa$ with dimension
of mass and $a, b \in \IC$ which are dimensionless parameters. The constant $\kappa$ has a particular r\^ole in the sense that it controls the 
separation of the points where the Yukawa couplings for the up and the down-type  quarks are generated, as we will now see.

This background for the Higgs field takes values in the subalgebra $\mathfrak{su}_{2}  \oplus \mathfrak{u}_{1} \oplus \mathfrak{u}_{1}$ orthogonal to $\mathfrak{su}_{5}^{\rm GUT}$ in $\mathfrak{e}_7$. As discussed in the previous section there are two possible assignments of matter fields that give rank one Yukawa couplings at tree-level. Here we recall the two possible assignments by specifying their charges under $SU(2) \times U(1) \times U(1)$
\begin{itemize}
\item[-] \textbf{Model A}
\be
\mathbf{10}_M : \mathbf{2}_{1,0}\,, \quad \mathbf{5}_U : \mathbf{1}_{-2,0}\,, \quad \bar{\mathbf{5}}_M : \mathbf{2}_{0,1}\,, \quad \bar{\mathbf{5}}_D : \mathbf{ 1}_{-1,-1}\,,
\ee
\item[-] \textbf{Model B}
\be
\mathbf{10}_M : \mathbf{2}_{1,0}\,, \quad \mathbf{5}_U : \mathbf{1}_{-2,0}\,, \quad \bar{\mathbf{5}}_M : \mathbf{ 1}_{-1,-1}\,, \quad \bar{\mathbf{5}}_D : \mathbf{2}_{0,1}\,.
\ee
\end{itemize}
These assignments specify how the Higgs field background (\ref{Phixy}) enters the zero mode equation for each matter fields, and therefore the curves at which they are localised. As in \cite{mrz15} we define $\Phi|_{{\mathcal R}_{\rm GUT}}$ as the action of $\langle \Phi_{xy} \rangle$ on the $\mathfrak{g}^\perp$ part of $(\CR_{\rm GUT}, \CR) \subset \textbf{133}$. For the model A we obtain
\be\label{eq:PhiA}\begin{split}
\Phi|_{\mathbf{10}_M} = \left(\begin{array}{c c}\mu_1^2 (a x -y )& m \\ m^2 x & \mu_1^2 (a x -y)\end{array}\right)\,, &\quad \Phi|_{\mathbf{5}_U} = -2\mu_1^2 (a x -y)\,,\\
 \Phi|_{\bar{\mathbf{5}}_M} = \left(\begin{array}{c c} \mu_2^2 (b x -y )+\kappa & m \\ m^2  x & \mu_2^2 (b x -y)+\kappa\end{array}\right)\,, &\quad \Phi|_{\bar{\mathbf 5}_D} = -\mu_1^2 (
a x - y ) - \mu_2^2 (b x -y)-\kappa\,.
\end{split}\ee
where the action in the model B may be easily obtained by simply interchanging the actions on $\bar{\mathbf 5}_M$ and $\bar {\mathbf 5}_D$. 
The location of the matter curve hosting the representation ${\mathcal R}_{\rm GUT}$ is then found by computing $\text{ det} \Phi|_{{\mathcal R}_{\rm GUT}} =0$. 
For the model A the explicit location is
\be\label{sigmas}\begin{split}
&\Sigma_{\mathbf{10}_M} : \mu_1^4 (a x -y)^2 - m^3 x =0\,, \quad \Sigma_{\bar{\mathbf 5}_M} : \left[\mu_2^2(b x - y )+\kappa\right]^2 - m^3 x = 0 \,,\\
&\Sigma_{\mathbf{5}_U} : \mu_1^2 (a x -y) = 0\,, \quad \Sigma_{\bar {\mathbf 5}_D} : \mu_1^2 (a x - y ) + \mu_2 ^2 (b x - y ) + \kappa =0\,.
\end{split}
\ee
This expression for the matter curves allows to compare the present model with the model of $E_8$ enhancement considered in \cite{mrz15}, see eq.(4.8) therein. In particular we see that the present model is more general, and that we recover the same matter curves as in \cite{mrz15} if we set $a=b$.\footnote{The precise dictionary connecting the two models would be
\be\label{eq:dict}\begin{split}
-(1+2d) \mu^2 &\raw \, \mu_2^2\\
-2d\kappa &\raw\,  \kappa
\end{split}\ee where the parameters in the left hand side of \eqref{eq:dict} are the ones appearing in \cite{mrz15} and the parameters in the right hand side are the ones of this paper. \label{dictionary}} However, as we will see below considering $a \neq b$ will be crucial in order to implement the doublet-triplet splitting mechanism of \cite{bhv2} and it will also greatly increase the region of parameters for which we can reproduce the empirical masses and mixing for charged MSSM fermions. 

Finally, Yukawa couplings for the matter fields are generated at the intersection of these matter curves. In particular the Yukawa coupling $\mathbf{10}_M\times \mathbf{10}_M\times \mathbf{5}_U$ of the up-type quarks is generated at the point where the curves $\Sigma_{\mathbf{10}_M}$ and $\Sigma_{\mathbf{5}_U}$ meet whereas the Yukawa coupling 
$\mathbf{10}_M \times \bar{\mathbf 5}_M \times \bar {\mathbf{5}}_D$ of the leptons and down-type quarks is generated where the curves $\Sigma_{\mathbf{10}_M}$, 
$\Sigma_{\bar {\mathbf 5}_M}$ and $\Sigma_{\bar {\mathbf 5}_D}$ meet. 
These two points are
\be\begin{split}
Y_U &: \Sigma_{\mathbf{10}_M}\, \cap \Sigma_{\mathbf{5}_U} = \{ x = y = 0\} = p_{\rm up}\,,\\
Y_{D/L} &: \Sigma_{\mathbf{10}_M}\, \cap \Sigma_{\bar{\mathbf{5}}_D} \cap \Sigma_{\bar{\mathbf 5}_M} = \{ x=x_0,\,  y = y_0\} = p_{\rm down}\,,
\end{split}\ee
where
\be\label{pdown}
x_0 = \frac{\kappa^2 \mu_1^4}{m^3 (\mu_1^2+\mu_2^2)^2}+ \mathcal O (\kappa^3)\,, \quad y_0 = \frac{\kappa}{\mu_1^2+\mu_2^2}\left(1+\frac{\kappa \mu_1^4
(a \mu_1^2 + b \mu_2^2)}{m^3 (\mu_1^2 +\mu_2^2)^2}\right)+\mathcal O (\kappa^3)\,.
\ee
This shows that the two Yukawa points do not necessarily coincide and that the parameter $\kappa$ controls the separation between them. Setting $\kappa = 0$ both couplings are 
generated at the same point while the separation of the two points increases with $\kappa$.

\subsubsection*{Real gauge}

The background fields described so far are in holomorphic gauge and therefore to achieve a physical solution, namely one in which the gauge fields are real and the D-term equations
are satisfied, it is necessary to go to a real gauge. This may be attained by simply performing a gauge transformation defined by an element $g$ of the complexified gauge
group $G_{\mathbb C}$ 
so that the D-term equations simply become a set of differential equations for $g$. More explicitly the effect of this gauge transformation on the background fields is the following
one
\be
\Phi_{xy} \rightarrow  g\, \Phi_{xy} \,g^{-1}\,, \quad A_{0,1} \rightarrow A_{0,1} + i g\, \bar \p g^{-1}\,,
\ee
where in our case we take $g \in SU(2)_{\mathbb C}$. Following \cite{cchv10} we take the following Ansatz for $g$
\be
g = \text{exp}\left[\frac{1}{2} f P\right]\,,
\ee
where $P = [E^+, E^-]$. After this gauge transformation the background fields become
\begin{subequations}
\begin{align}
\Phi_{xy} =&\, m\left(e^f E^++mxe^{-f}E^-\right)+{\mu}_1^2\left({a}x-y\right)Q_1+\left[{\mu}_2^2\left({b}x-y\right)+\kappa\right]Q_2\,,\\
A_{0,1} = & -\frac{i}{2}\bar \p f P\,.
\label{primitiveA}
\end{align}
\end{subequations}
Plugging this Ansatz into the D-term equations one obtains a differential equation for $f$.
More precisely, by taking the following expression for the K\"ahler form
\be
\omega = \frac{i}{2} (d x \wedge d \bar x + d y \wedge d \bar y)\,,
\ee
the D-term equations become
\be\label{eq:D2}
( \p_x \bar \p_{\bar x} + \p_y \bar \p_{\bar y} ) f = m^2 (e^{2f} -m^2 |x|^2 e^{-2f})\,.
\ee
As in \cite{fmrz13,  mrz15} we take $f$ to depend only on $r = (x \bar x)^{\frac{1}{2}}$. Defining $s = \frac{8}{3} (m r)^{\frac{3}{2}}$ and $h$ as
\be
e^{2f} = m r e^{2 h}\,,
\ee 
eq.\eqref{eq:D2} becomes
\be
\left(\frac{d^2}{ds^2}+\frac{1}{s}\frac{d}{ds}\right) h = \frac{1}{2} \sinh (2h)\,.
\ee
This is a particular instance of the Painlev\'e III equation and its solution valid over the entire complex plane may be found in \cite{mtw77}. Since our model is defined
only in a local patch of $S_{GUT}$ it suffices to expand the solution around the origin and retain only the lowest order terms in $r$
\be
f (r ) = \log c + c^2 m^2 r^2 + m^4 r^4 \left(\frac{c^4}{2}-\frac{1}{4 c^2}\right)+\dots \,.
\ee
The constant $c$ in this equation may be fixed if we ask for a solution regular for all values of $r$, and the explicit value is
\be\label{eq:cvalue}
c = {3}^{\frac{1}{3}} \frac{\Gamma \left[\frac{2}{3}\right]}{\Gamma \left[\frac{1}{3}\right]}\sim 0.73\,.
\ee
However since we are only interested in a local solution we shall not restrict to this value in the following and leave $c$ as a free parameter controlling the strength of the non-primitive
flux at the origin.

\subsection{Primitive fluxes}\label{sec:flux}

While the background fields specified in the previous section are a consistent solution to the equations of motion it is still possible to consider a more general supersymmetric background for the gauge field strength $F$. In particular one may add an extra flux besides (\ref{primitiveA}) that is primitive and commutes with $\Phi$. The most general choice of gauge flux that satisfies these constraints and does not break $SU(5)_{GUT}$ is
\be
F_Q = i (dx \wedge d\bar x- dy \wedge d \bar y)\left[M_1 Q_1 +M_2 Q_2 \right]+i (d x \wedge d\bar y +dy \wedge d\bar x)\left[N_1 Q_1 +N_2 Q_2\right]\,.
\ee
This flux has the main effect of inducing 4d chirality in the matter field spectrum because modes of opposite chirality will feel it differently. We will discuss more in 
detail how the presence of fluxes selects a preferred 4d chirality later in this section.

Finally, an important ingredient missing so far is a mechanism to achieve the breaking of $SU(5)_{GUT}$ down to the SM gauge group. We choose to employ the standard mechanism 
for GUT breaking in F-theory \cite{dw2,bhv2} and add a flux along the hypercharge generator. We assume that the integrals for the hypercharge flux are such that no mass term is generated for the hypercharge gauge boson, a condition that can only be checked in a global realisation of our model. In our local approach we may choose the following parametrisation for this flux
\be
F_Y = i \left[\tilde N_Y (dy \wedge d \bar y-dx \wedge d\bar x)+N_Y (dx \wedge d\bar y + dy \wedge d\bar x)\right]Q_Y\,,
\ee
where we defined the hypercharge generator as
\be
Q_Y = \frac{1}{3}\left(H_1 + H_2 +H_3\right)-\frac{1}{2}(H_4 + H_5)\,.
\ee
To summarise, the total primitive flux present in our model is
\be
F_p = i Q_R (dy \wedge d \bar y - dx \wedge d\bar x)+i Q_S (dy \wedge d \bar x + dx \wedge d\bar y)\,,
\ee
where we defined the generators
\be\label{qrqs}
Q_R = - M_1 Q_1 - M_2 Q_2 +\tilde N_Y Q_Y\,,\quad  Q_S = N_1 Q_1+N_2 Q_2 +N_Y Q_Y\,.
\ee
These fluxes will enter explicitly in the equations of motion for the physical zero modes and because of this they will enter directly in the expression of the Yukawa couplings. Just like in \cite{afim11,fimr12,fmrz13,mrz15} the holomorphic Yukawa couplings that enter in the superpotential are not affected by the fluxes. However, the physical Yukawa couplings will depend on them after imposing correct normalisation of the kinetic terms for the matter fields. We have chosen to summarise how the primitive flux is felt by the various MSSM fields for the case of the model A in Table \ref{t:sectors} specifying the
two combinations $q_R$ and $q_S$ that will be relevant for the computation in the following sections. 

\begin{table}[htb]
\renewcommand{\arraystretch}{1.2}
\setlength{\tabcolsep}{2pt}
\begin{center}
\begin{tabular}{|c||c|c|l||c|c|}
\hline
MSSM & Sector & $SU(2) \times U(1) \times U(1)$ & $G_{\rm MSSM}$  & $q_R$ & $q_S$\\
\hline
\hline
$Q$&$\mathbf{10}_M$&$\mathbf{2}_{1,0}$&$(\mathbf{3},\mathbf{2})_{-\frac{1}{6}}$&$-\frac{1}{6}\tilde N_Y-M_1$&$-\frac{1}{6} N_Y+N_1$\\
\hline
$U$&$\mathbf{10}_M$&$\mathbf{2}_{1,0}$&$(\mathbf{\bar 3},\mathbf{1})_{\frac{2}{3}}$&$\frac{2}{3}\tilde N_Y-M_1$&$\frac{2}{3} N_Y+N_1$\\
\hline 
$E$&$\mathbf{10}_M$&$\mathbf{2}_{1,0}$&$(\mathbf{1},\mathbf{1})_{-1}$&$-\tilde N_Y-M_1$&$- N_Y+N_1$\\
\hline
$D$&$\mathbf{\bar 5}_M$&$\mathbf{2}_{0,1}$&$(\mathbf{\bar 3},\mathbf{1})_{-\frac{1}{3}}$&$-\frac{1}{3}\tilde N_Y-M_2$&$-\frac{1}{3} N_Y+N_2$\\
\hline
$L$&$\mathbf{\bar 5}_M$&$\mathbf{2}_{0,1}$&$(\mathbf{1},\mathbf{2})_{\frac{1}{2}}$&$\frac{1}{2}\tilde N_Y-M_2$&$\frac{1}{2} N_Y+N_2$\\
\hline
$H_u$&$\mathbf{5}_U$&$\mathbf{1}_{-2,0}$&$(\mathbf{1},\mathbf{2})_{-\frac{1}{2}}$&$-\frac{1}{2}\tilde N_Y+2M_1$&$-\frac{1}{2} N_Y-2N_1$\\
\hline
$H_d$&$\mathbf{\bar 5}_D$&$\mathbf{1}_{-1,-1}$&$(\mathbf{1},\mathbf{2})_{\frac{1}{2}}$&$\frac{1}{2}\tilde N_Y+M_1+M_2$&$\frac{1}{2} N_Y-N_1-N_2$\\
\hline
\end{tabular}
\end{center}
\caption{Different sectors and charges for the $E_7$ model of this section. Here $q_R$ and $q_S$ are the $E_7$ operators (\ref{qrqs}) evaluated at each different sector. All the multiplets in the table have the same chirality.}
\label{t:sectors}
\end{table}
\subsubsection*{Local chirality of matter fields}

One of the most important consequences of the addition of gauge fluxes on the worldvolume of 7-branes is the generation of a chiral spectrum in the 4d effective theory.
It is possible to compute the net chiral spectrum of the modes localised on a matter curve $\Sigma$ as an index \cite{bhv1}
\be
\chi\left(\Sigma,\mathcal L \otimes K_{\Sigma}^{\frac{1}{2}}\right) = \int_{\Sigma} c_1 (\mathcal L)\,,
\ee
where $\mathcal L$ is a line bundle on $\Sigma$ whose first Chern class is equal to the magnetic flux threading the matter curve. Therefore a suitable choice of fluxes can
give the correct chiral spectrum in the 4d theory. Moreover, since part of the flux triggers the breaking of the GUT group, fields in different representations of the SM group that are
in the same representation of the GUT group may have a different chiral spectrum in 4d. This kind of mechanism allows for a simple implementation of doublet-triplet splitting in F-theory
GUTs by imposing the absence of massless Higgs triplets in the 4d theory. 

Notice that in our local setup we are not able to compute explicitly the chiral index for the various matter representations because this would require to specify the geometry around a patch containing $S_{GUT}$ and in particular the matter curves $\Sigma$. It is however still possible to discuss chirality in our local model by employing the concept of local chirality. This notion introduced in \cite{Palti:2012aa} amounts to compute a chiral index for those wavefunctions which are localised around the Yukawa point. To gain a better understanding of how local chirality is formulated it is useful to consider models of magnetised D9-branes which are T-dual to our setting, as in \cite{afim11}. In order to do so we identify the gauge connection $A_{\bar z}$ with $\Phi$ where we called $z$ the direction transverse to the 7-branes. All fields do not depend on $z$ and therefore $F_{x \bar z} = D_x \Phi$ and $F_{y \bar z} = D
_y \Phi$ and so on. To formulate local chirality we need the expression of the index of the Dirac operator which for a representation $\mathcal R$ is
\be\label{eq:locind}
{\rm index}_{\mathcal R}  \slashed D=\frac{1}{48(2\pi)^2}\int \left ( \tr_{\mathcal R}\,F\,\w\, F\,\w \,F-\frac{1}{8}\tr_{\mathcal R}\,F\,\w\, \tr R\,\w\, R \right )\,.
\ee
Asking for the existence of a chiral mode in the representation $\mathcal R$ amounts to the condition $\mathcal I_{\mathcal R} < 0$ where $\mathcal I_{
\mathcal R}$ is the integrand in \eqref{eq:locind}. Note that since $\mathcal I_{\mathcal R} = - \mathcal I_{\overline{\mathcal R}}$ the spectrum in the 4d theory will 
be chiral. Taking a local patch where we can approximate our configuration by constant fluxes and vanishing curvature we find
\begin{eqnarray}\label{indR}
\mathcal I_{\mathcal R}\equiv\frac{i}{6}\tr_{\mathcal R}\,\left (F\,\w\, F\,\w \,F\right )_{x\bar xy\bar yz\bar z}=i\,\tr_{\mathcal R}\big ( F_{x\bar x}\{F_{y\bar y},F_{z\bar z}\}+F_{x\bar z}\{F_{y\bar x},F_{z\bar y}\}+ \\\nonumber
F_{x\bar y}\{F_{y\bar z},F_{z\bar x}\}- \{F_{x\bar x},F_{y\bar z}\}F_{z\bar y} - \{F_{x\bar y},F_{y\bar x}\}F_{z\bar z} - \{F_{x\bar z},F_{y\bar y}\}F_{z\bar x}  \big ).
\end{eqnarray}
Then, evaluating this expression for the various sectors of our model we obtain\footnote{In writing $\mathcal I _{\mathbf{10,2}}$ and
$\mathcal I _{\mathbf{\bar 5,2}}$ we have neglected some terms involving $\mu_1$ and $\mu_2$. We chose to do so because as we will discuss later 
we shall restrict to the case $\mu_1,\mu_2 \ll m$ implying that these additional terms will give negligible contributions to the local chiral index.}
\bea
\mathcal I _{\mathbf{10,2}} &= & -2 m^4 c^4 q_R ^{(\mathbf{10,2})}\\
\mathcal I _{\mathbf{\bar 5,2}} &= & -2 m^4 c^4 q_R ^{(\mathbf{\bar 5,2})}\\
\mathcal I _{\mathbf{ 5,1}} &= & -4 \mu_1^4[q_R ^{(\mathbf{ 5,1})} (|a|^2-1) +2 \text{Re}[a] q_S^{(\mathbf{ 5,1})} ]\\
\mathcal I _{\mathbf{\bar  5,1}} &= & -
\left\{q_R ^{(\mathbf{ \bar 5,1})}[|a \mu_1^2 +b \mu_2^2|^2 - |\mu_1^2+\mu_2^2|^2 ]+2 q_S^{(\mathbf{\bar  5,1})}\text{Re}[(a\mu_1^2 +b\mu_2^2)(\mu_1^2+\mu_2^2)]\right\}\,.
\eea

The conditions that we need to impose in order to obtain the correct chiral spectrum in 4d are the following ones
\be\begin{split}\label{eq:chirals}
\mathcal I_{\mathcal R} < 0 &\,, \qquad \mathcal R = Q,U,E,D,L,H_u,H_d\,,\\
\mathcal I_{\mathcal R} = 0 &\,, \qquad \mathcal R = T_u, T_d\,.
\end{split}\ee
We spell out the explicit form of these conditions for our models in Appendix \ref{ap:chiral}, where we also write the explicit form of the equations \eqref{eq:chirals} and discuss
the existence of solutions to the system. As shown in there for the particular case of $a=b=1$ considered in \cite{mrz15} the previous system does not admit solutions, and therefore it is 
not possible (at least in terms of local chirality) to obtain the MSSM chiral spectrum without Higgs triplets. Therefore we are led to consider models where $a \neq b$ and so, compared to the analysis in \cite{mrz15} our configurations have one further parameter $(a-b)$. As we will see in section \ref{s:fitting} imposing that this parameter is non-vanishing will allow to fit the empiric data for fermion masses in a much wider region of parameter space.

\subsection{Residue formula for Yukawa couplings}
\label{sec:residuemain}

Knowing the distribution of matter fields on each matter curve it is possible to compute the holomorphic Yukawa couplings by simply performing a dimensional reduction of the 7-brane superpotential
\be\label{eq:superp}
W = m_*^4\int_S \text{Tr}\left(\Phi \wedge F\right) +\frac{\eps}{2}\theta_0 \text{Tr}\left(F \wedge F\right)\,.
\ee
As discussed in section \ref{s:yukawas}, $\theta_0$ is a holomorphic section of a line bundle on $S$ and
$\epsilon$ is a parameter that measures the strength of the non-perturbative effect. It is important to note that this additional term in the superpotential will affect the supersymmetry
equations for the background that we discussed in the previous sections. This implies that the background values of $\Phi$ and $F$ will be deformed and have $\CO(\eps)$ 
corrections. As shown in \cite{fimr12} this does not affect the computation of holomorphic Yukawa couplings and so we may safely ignore this background shift in the discussion below.

To obtain the zero mode equations of motion we separate the 7-brane fields into background and fluctuations around the background
\be\label{expan}
\Phi = \langle \Phi \rangle + \varphi\,, \quad A= \langle A \rangle + a\,,
\ee
and retain only the terms linear in the fluctuations in the supersymmetry equations obtained from the superpotential in (\ref{eq:superp}). The resulting zero mode equations are
\be\label{eq:fluctF}\begin{split}
\bar \p_{\langle A\rangle} a &=0\,,\\
\bar \p_{\langle A\rangle} \varphi &=  i [a, \langle \Phi\rangle ]- \eps \p \theta_0 \wedge (\p_{\langle A\rangle} a +\bar \p_{\langle A\rangle} a^\dagger)\,.
\end{split}
\ee
where we have taken into account the shift in the value of $\langle \Phi \rangle$ as compared to (\ref{Phixy}) due to non-perturbative corrections, see  \cite{fimr12,fmrz13} for details.
The same procedure  applied to the D-term equation will yield an additional equation for the zero modes but we shall neglect it in this section, for we are only interested  
in the holomorphic part of the Yukawa couplings. The solution of the system (\ref{eq:fluctF}) is 
\be\label{eq:Fsol}\begin{split}
a &= \bar \p _{\langle A \rangle} \xi \,,\\
\varphi  &=h-i [\langle \Phi\rangle,\xi] +\eps \p \theta_0 \wedge (a^\dag - \p_{\langle A\rangle} \xi)\,,
\end{split}
\ee
where $\xi$ is a section of $\Omega^{(0,0)}(S) \otimes \text{ad}(E_7)$ and $h$ is a holomorphic section of $\Omega^{(2,0)}(S) \otimes \text{ad}(E_7)$. 
The presence of terms involving $a^\dag$ in  (\ref{eq:Fsol})  is a bit puzzling at first sight because it seems that non-holomorphic terms may be present in the 4d 
superpotential. However when performing the dimensional reduction of the 7-brane superpotential these terms will appear only in total derivatives and will therefore be
absent in the 4d superpotential \cite{fimr12}. Indeed, plugging the solutions (\ref{eq:Fsol}) into the superpotential and evaluating cubic terms in the fluctuations one finds that the Yukawa couplings read \cite{cchv09,fimr12,cchv10,fmrz13} 
\be
Y = -i \frac{m_*^4}{3} \int_S \text{Tr} (h \wedge \bar \p_{\langle A \rangle} \xi \wedge \bar \p_{\langle A \rangle} \xi)\,.
\ee
Finally, it is interesting to note that the computation of Yukawa couplings can be translated in a simple residue computation, as first noticed in \cite{cchv09} and generalised in \cite{fimr12,fmrz13} for the setup at hand. The final expression reads
\be
Y =  m_*^4 \pi^2 f_{abc}\,\text{Res}_p \left[\eta^a \eta^b h_{xy}\right] =  m_*^4 \pi^2 f_{abc} \int_{\mathcal{C}}\eta^a \eta^b h_{xy}d x \wedge dy
\label{yukres}
\ee
with $\mathcal{C}$ a cycle in $\mathbb{C}^2$ which can be continuously contracted to a product of unit circles surrounding the Yukawa point $p$ without encountering singularities in the integrand. Also the function $\eta$ is defined as
\be
\eta = - i \Phi^{-1}\left[h_{xy}+i \epsilon\p_x \theta_0 \p_y \left (\Phi^{-1} h_{xy}\right)-i \epsilon\p_y \theta_0 \p_x \left (\Phi^{-1} h_{xy}\right)\right]\,.
\label{eta}
\ee

\subsection{Holomorphic Yukawa couplings for the $E_7$ model}

We now have all the necessary ingredients to perform the computation of the Yukawa couplings in both $E_7$ models. Here we will only report the results for the model A deferring the
results for the model B to Appendix \ref{ap:modelB}. We focus our attention on the matter curves including the fields charged under the MSSM gauge group and therefore to the two Yukawa matrices for the couplings $\mathbf{10}_M \times \mathbf{10}_M \times \mathbf{5}_U$ and $\mathbf{10}_M \times \mathbf{\bar 5}_M \times \mathbf{\bar 5}_D$. The functions $h_{xy}$ for the different fields are
\be\label{haches}
\begin{array}{ll}
h_{\mathbf{10}_M} = \g_{10,i}\,m_*^{3-i}(ax-y)^{3-i}& h_{\mathbf{ \bar 5}_M}=  \g_{5,i}\,m_*^{3-i}(a(x-x_0)-(y-y_0))^{3-i}\\
h_{\mathbf{5}_U} = \gamma_U & h_{\mathbf{\bar 5}_D} = \gamma_D,
\end{array}
\ee
where $(x_0,y_0)$ corresponds to the coordinates (\ref{pdown}) of the down-type Yukawa point $p_{\rm down}$, while recall that $p_{\rm up}$ is located at the origin. Finally, the constants $\g_{10,i}, \g_{5,i},\g_U,\g_D$ are normalisation factors to be computed in the next section and $i=1,2,3$ is a family index. 
With this form one can compute the functions $\eta$ in (\ref{eta}) which in turn are needed to compute the holomorphic couplings via the residue formula (\ref{yukres}). We relegate the expressions for such $\eta$'s to Appendix \ref{ap:holoyuk} and turn to discuss the Yukawa matrices that result from them. 

Below we display the Yukawa matrix for the up-type quarks up to first order in the expansion parameter $\eps$. For the Yukawa matrix of down
quarks and leptons we find an explicit dependance on $\kappa$, the parameter controlling the separation between the two Yukawa points. Since the 
dimensionless combination $\tilde \kappa = \kappa / m_*$ will turn out to be very small we chose to retain only the first two orders in $\tilde \kappa$
in the Yukawa matrix (dropping also terms of order $\CO( \eps \tilde \kappa)$ which are extremely suppressed).
Moreover for the Yukawa matrix of down quarks and leptons we also perform an expansion on the parameter $(a-b)$ which we will eventually find to be small as well. Our computations in section \ref{s:fitting} will however be based on the full $(a-b)$ dependence of $Y_{D/L}$, which can be found in appendix \ref{app:YukA}.
\be\label{hyukU}
Y_U =  \frac{\pi^2\,\g_U\,\g_{10,3}^2}{2\rho_m\rho_\mu}\left (\begin{array}{ccc}
0&0&\tilde \eps\frac{\g_{10,1}}{2\rho_\mu\g_{10,3}}\\
0&\tilde \eps\frac{\g_{10,2}^2}{2\rho_\mu\g_{10,3}^2}&0\\
\tilde \eps\frac{\g_{10,1}}{2\rho_\mu\g_{10,3}}&0&1\end{array}\right )+ \mathcal{O} (\eps^2)\,,
\ee
\be\label{hyukD}
Y_{D/L} = Y_{D/L}^{(0)} + (a-b) \,Y_{D/L}^{(1)}+ \CO((a-b)^2)
\ee
where
\be
Y^{(0)}_{D/L} =-\frac{\pi ^2 \gamma _{5,3} \gamma _{10,3} \gamma _D}{(d+1)
   \rho _{\mu } \rho _m} \left(
\begin{array}{ccc}
 0 & \tilde \kappa  \tilde  \epsilon\frac{2   \gamma _{5,2} \gamma _{10,1}
 }{(d+1)^2 \rho _{\mu }^2\gamma
   _{5,3} \gamma _{10,3} } & \frac{\g_{10,1}}{(d+1) \rho_\mu \g_{10,3}}\left(\frac{2 \tilde \kappa^2}{(d+1)^2\rho_\mu}-\tilde \eps\right)\\
  \tilde \kappa \tilde  \eps\frac{   \gamma _{5,1} \gamma
   _{10,2} }{(d+1)^2\rho _{\mu }^2
   \gamma _{5,3} \gamma _{10,3} } & -\tilde  \eps\frac{  \gamma _{5,2} \gamma _{10,2}}{(d+1) \rho _{\mu
   } \gamma _{5,3} \gamma _{10,3}} & -\tilde{\kappa }\frac{ \gamma _{10,2}}{(d+1)    \rho _{\mu }\gamma _{10,3}
} \\
\tilde \eps \frac{  
   \gamma _{5,1}}{(d+1) \rho _{\mu } \gamma _{5,3}} & 0 & 1
   \\
\end{array}
\right)
\ee
\be
Y_{D/L}^{(1)}= - \frac{\pi ^2}{(d+1)^3} \left(\begin{array}{ccc}0 & y^{(12)} & y^{(13)} \\  y^{(21)} & y^{(22)} & y^{(23)}\\y^{(31)} & y^{(32)} &y^{(33)}\end{array}\right)
\ee
with the entries given by
\be
y^{(12)} = - \eps \tilde \kappa \, \frac{(d-1) \gamma _{5,2} \gamma _{10,1} \gamma _D
   \theta _y}{(d+1) \rho _{\mu }^3 \rho _m}
\ee
\be
y^{(13)} = \frac{\left(d-1\right) \gamma _{5,3} \gamma
   _{10,1} \gamma _D}{2  \rho _{\mu }^2 \rho _m}\left[\epsilon  \,\theta _y-\tilde \eps \,\tilde \kappa\frac{4 d (5 d-1) \rho _{\mu }
   }{\left(d^2-1\right)
   \rho _m^{3/2}}\right]
   \ee
\be
y^{(21)} =- \eps \tilde \kappa\, \frac{ (d-1) \gamma _{5,1} \gamma _{10,2} \gamma _D
   \theta _y}{2 (d+1) \rho _{\mu }^3 \rho _m}
\ee
\be
y^{(22)} = \frac{\left(d-1\right) \gamma _{5,2} \gamma
   _{10,2} \gamma _D}{2 \rho _{\mu }^2 \rho _m}\left[\eps \,\th_y+\tilde \eps \tilde  \kappa\,\frac{18 d\rho _{\mu } }{\left(d^2-1\right) \rho
   _m^{3/2}}\right]
\ee
\be
y^{(23)} = \frac{3 d^2 \gamma _{5,3} \gamma _{10,2} \gamma
   _D}{ \rho _m^{5/2}}\left[\tilde \eps +\tilde{\kappa }^2\frac{2 }{d(1+d) \rho _{\mu }}\right]
\ee
\be
y^{(31)} = \frac{ (d-1) \gamma _{5,1} \gamma _{10,3} \gamma
   _D}{2  \rho _{\mu }^2 \rho _m}\left[\eps \, \th_y +\tilde \kappa \tilde \eps\frac{4 (d-2)   \rho _{\mu }
 }{\left(d^2-1\right)
   \rho _m^{3/2}}\right] 
\ee
\be
y^{(32)} = - \tilde \eps \,\frac{3  d   \gamma _{5,2} \gamma _{10,3}
   \gamma _D}{
   \rho _m^{5/2}}
\ee
\be
y^{(33)} = - \tilde{\kappa } \,\frac{2 d \gamma _{5,3} \gamma
   _{10,3} \gamma _D}{ \rho _m^{5/2}}
\ee
and where we have defined the following quantities
\be
d = \frac{\mu^2_2}{\mu_1^2}\,, \quad \rho_\mu = \frac{\mu_1^2}{m_*^2}\,, \quad \rho_m = \frac{m^2}{m_*^2}\,, \quad \tilde \kappa = 
\frac{\kappa}{m_*}\,, \quad \tilde \eps = \eps (\th_x + a \th_y)\,.
\ee

\section{Normalisation factors and physical Yukawas}
\label{s:norm}


So far we have been performing the computation of the Yukawa couplings merely at the holomorphic level, i.e. we have
performed the computation of the four dimensional superpotential for the zero modes. To complete the computation and obtain
results comparable with measured data it is necessary to compute the kinetic terms of the zero modes and take them to a basis
where they are canonically normalised.  To compute the kinetic terms it is necessary first to go in a real gauge and solve the zero mode equations in there, which has the 
effect to induce a dependance on the local flux densities in the kinetic terms.

In this section we will solve the wavefunctions in a real gauge and use this result to obtain the various normalisation factors. In the 
sectors affected by the T-brane background we will not be able to find an analytical solution. However like in \cite{fmrz13,mrz15}
we will be able to find an approximate solution in some regions of the parameter space of our local model. We will first compute
the wavefunctions that correspond to the tree-level superpotential and show that no kinetic mixing is present at the level of approximation that we are working. We will then 
include the non-perturbative corrections and argue that the result does not change. 

To summarise, in this section we will compute the normalisation factors for the chiral wavefunctions of the $E_7$ model. At tree-level and $\CO(\eps)$ they correspond to kinetic terms with a diagonal structure, a result that is not changed by non-perturbative effects. This implies that we may compute the final result for the physical
Yukawa couplings by employing the holomorphic result discussed in the previous section together with the normalisation factors
that we are going to derive below.

\subsection{Perturbative wavefunctions}

Without non-perturbative corrections the equations of motion for the zero modes can be obtained from \eqref{Fterm7} expanding
the fields as $\Phi = \langle \Phi\rangle + \varphi$ and $A = \langle A \rangle + a$ and retaining only the terms linear in
$\varphi$ and $a$. The resulting equations are
\begin{subequations}\label{zeropert}
\begin{align}
\bar \p_{\langle A\rangle} a=&\,0\,,\label{treeF1}\\
\bar \p_{\langle A\rangle} \varphi =&\,i [a , \langle \Phi\rangle]\,,\label{treeF2}\\
\omega \wedge  \p_{\langle A\rangle} a=&\,\frac{1}{2}[\langle \bar \Phi \rangle,\varphi]\,.\label{treeD}
\end{align}
\end{subequations}
In \eqref{zeropert} we choose $\langle \Phi\rangle$ and $\langle A\rangle$ to be background fields in a real gauge. 
These equations may be solved by using techniques already employed in \cite{afim11,fimr12,fmrz13,mrz15}. To keep the discussion contained
we will simply quote the results in this section deferring more details regarding the computation to Appendix \ref{ap:holoyuk}.

Henceforth we are going to use the following notation for the zero modes
\be
\vec \varphi_\rho=  \left(\begin{array}{c}
a^s_{\bar x}\\
a^s_{\bar y}\\
\varphi^s_{xy}\end{array}\right) E_{\rho,s}
\ee
where $E_{\rho,s}$ denotes the particular set of roots, labelled by $s$, corresponding to a given sector  $\rho$. In our models 
we have that for the sectors unaffected by the T-brane background $s$ takes a single value whereas in the other sectors 
we have that $s$ takes two values.

\subsubsection*{Sectors not affected by T-brane}

In the sectors not affected by the T-brane background the solution may be computed analytically. In the models we 
consider we have two sectors that fall in this class and transform as $(\mathbf{5},\mathbf 1)_{-2,0}$ and $(\mathbf{5},\mathbf 1)_{1,1}$ under
$SU(5) \times SU(2) \times U(1) \times U(1)$. We recall here that in both models $\mathbf 5_U :=(\mathbf{5},\mathbf 1)_{-2,0}$,
while $\mathbf{\bar 5}_D:=(\mathbf{\bar 5},\mathbf 1)_{-1,-1}$ in the model A and $\mathbf{\bar 5}_M:=(\mathbf{\bar 5},
\mathbf 1)_{-1,-1}$ in the model B.

The solution for both sectors is the following one 
\be
\vec \varphi = \left(\begin{array}{c}-\frac{i \zeta}{2 \tilde \mu_a} \\ \frac{i (\zeta-\lambda)}{2 \tilde \mu_b}\\1\end{array}\right) \chi (x,y) 
\ee
where 
\be
\chi (x,y) = e^{\frac{q_R}{2} (x \bar x -y \bar y)-q_S\text{Re} (x \bar y)+ (\tilde \mu_a x+ \tilde \mu_b y)(\zeta_1 \bar x - \zeta_2 \bar y)}\,f(\zeta_2x+\zeta_1y)
\ee
and where we have defined 
\be
\zeta = \frac{\tilde \mu _a \left(4 \tilde \mu _a\tilde  \mu _b+\lambda  q_S\right)}{\tilde \mu _a q_S+\tilde \mu _b \left(\lambda +q_R\right)}\,,\quad 
\zeta_1 = \frac{\zeta}{\tilde \mu_a} \,, \quad \zeta_2 = \frac{\zeta-\lambda}{\tilde \mu_b}\,,
\ee
and $\lambda$ is the lowest solution to the cubic equation \eqref{eq:cub1}. The parameters $\tilde \mu_a $ and $\tilde \mu_b$ are
directly related to the ones describing the background Higgs fields and for both sectors are given by
\begin{center}\begin{tabular}{|c |c |c|}
\hline
 & $(\mathbf{5},\mathbf 1)_{-2,0} $& $(\mathbf{\bar 5},\mathbf 1)_{-1,-1}$\\
 \hline 
 $\tilde \mu_a $ & $ a \mu_1^2$ & $\frac{1}{2}(a \mu_1^2 + b \mu_2^2 )$\\
 \hline
 $\tilde \mu_b$ &  $-\mu_1^2$ & $-\frac{1}{2}( \mu_1^2 +  \mu_2^2 )$\\
 \hline
\end{tabular}\end{center}
Finally, the function $f(\zeta_2 x + \zeta_1 y)$ is a holomorphic function which can be approximated by a constant if the sector we consider contains
an MSSM Higgs. In the remaining case, namely the identification $\mathbf{\bar 5}_M:=(\mathbf{\bar 5}, \mathbf 1)_{-1,-1}$ for the model B we can choose 
\be
f_{\bar{\mathbf{5}}_M }^i (x,y ) = m_*^{3-i } (\zeta_2 (x-x_0) + \zeta_1 (y-y_0))^{3-i}\,,
\ee
where $i=1,2,3$ is a family index.
\subsubsection*{Sectors affected by T-brane}

In the two sectors affected by the T-brane background the equations of motion become more complicated. Here the fields
involved in the solution are doublets of $SU(2)$ in the decomposition of $E_7$ as $SU(5) \times SU(2) \times U(1) \times U(1)$
and therefore we are going to write the solution as
\be
\vec \varphi = \left(\begin{array}{c} a^+_{\bar x}\\ a^+_{\bar y}\\\varphi^+ _{xy}\end{array}\right) E^++\left(\begin{array}{c} a^-_{\bar x}\\ a^-_{\bar y}\\\varphi^- _{xy}\end{array}\right) E^-=
\vec \varphi_+E^+ +\vec \varphi_- E^-\,,
\ee
where we denote with a + the upper component of the $SU(2)$ doublet and with a $-$ the lower one. The equations for the 
zero modes are generally difficult to solve analytically. Nevertheless as discussed in appendix \ref{ap:holoyuk} in the limit $\mu_1,\mu_2,\kappa \ll m$ it is possible to
find approximate solutions. In both models the solution for the $\mathbf {10}_M$ sector in real gauge is
\be
\vec \varphi_{10}^i = \gamma_{10}^i \left(\begin{array}{c}\frac{i \lambda_{10}}{m^2}\\-i\frac{\lambda_{10} \zeta_{10}}{m^2}\\0\end{array}\right)e^{f/2} \chi_{10}^i E^++\gamma_{10}^i \left(\begin{array}{c}0\\0\\1\end{array}\right)e^{-f/2} \chi_{10}^i E^-
\ee
with $\lambda_{10}$ the negative solution to the cubic equation \eqref{cub10} and $\zeta_{10} = -q_S / (\lambda_{10} -q_R)$. Finally the wavefunctions $\chi_{10}^i$ are 
\be
\chi_{10}^i = e^{\frac{q_R}{2}(|x|^2-|y|^2)-q_S (x\bar y+y\bar x)+\lambda_{10} x(\bar x- \zeta_{10} \bar y)} g_{10}^i(y+\zeta_{10} x)\,,
\ee
where $g_{10}^i$ are holomorphic functions of the variable $y+\zeta_{10} x$ and $i=1,2,3$ is a family index. As in \cite{fimr12,fmrz13} we choose these holomorphic functions in the following way
\be\label{fam10}
g_{10}^i (y+\zeta_{10} x) = m_*^{3-i} (y+\zeta_{10} x)^{3-i}\,.
\ee
The other sector affected by the T-brane background is the $(\mathbf{\bar 5},\mathbf{2})_{0,1}$. In the model A we identify
it with the $\bar {\mathbf 5}_M$ sector and the solution is
\be
\vec \varphi_{5}^i = \gamma_{5}^i \left(\begin{array}{c}\frac{i \lambda_{5}}{m^2}\\-i\frac{\lambda_{5} \zeta_{5}}{m^2}\\0\end{array}\right)e^{i\tilde\psi +f/2} \chi_{5}^i(x,y-\nu/a) E^++\gamma_5^i \left(\begin{array}{c}0\\0\\1\end{array}\right)e^{i\tilde\psi-f/2} \chi_{5}^i(x,y-\nu/a) E^-
\ee
with $\tilde\psi$ defined in \eqref{gaugepar2} and $\nu=\kappa/\mu_2^2$. Also, $\lambda_{5}$ is defined as the lowest solution to (\ref{cub10}) and $\zeta_{5} = -q_S / (\lambda_{5} -q_R)$. Finally the wavefunctions $\chi_{5}^i$ are 
\be
\chi_{5}^i(x,y) = e^{\frac{q_R}{2}(|x|^2-|y|^2)-q_S (x\bar y+y\bar x)+\lambda_{5} x(\bar x- \zeta_{5} \bar y)} g_{5}^i(y+\zeta_{5} x)\,,
\ee
where $g_{5}^i$ are holomorphic functions of $y+\zeta_{5} x$ and $i=1,2,3$ is a family index. Analogously, the family functions are
\be\label{fam5}
g_{5}^i (y+\zeta_{5} x) = m_*^{3-i} (y+\zeta_{5} x)^{3-i}\,.
\ee
In the model B where we identify the $(\mathbf{\bar 5},\mathbf{2})_{0,1}$ sector with the $\bar {\mathbf 5}_D$ we find exactly
the same solution and the only difference involves the function $g_5(y+ \zeta_5 x)$ which in this sector is taken to be constant.

\subsection{Normalisation factors}

With the information regarding the perturbative wavefunctions we can compute the normalisation factors for the various 
sectors. The factor appearing in front of the kinetic terms for the various fields is
\be
K_\rho^{ij}\, =\, \langle \vec{\vphi}_{\rho}^{i} | \vec{\vphi}_{\rho}^{j} \rangle \, = \,
m_*^4 \int_S \tr \,( \vec{\vphi}_{\rho}^{i}{}^\dag \cdot \vec{\vphi}_{\rho}^{j})\, {\rm d vol}_S
\label{4dnorm}
\ee
which follows from direct dimensional reduction.

For both models we find that $K^{ij}_\rho =0$ for $i\neq j$ and  no kinetic mixing is present. Therefore the choice of the normalisation factors $|\g_\rho^i|^2 = (K^{ii}_\rho)^{-1}$
is sufficient to ensure canonically normalised kinetic terms. The computation of these factors is similar to the one performed in \cite{fimr12,fmrz13} and will not be repeated it here. For the model A we find
\begin{subequations}\label{normA}
\begin{align}
|\gamma_{U/D}|^2=&-\frac{4^{2}}{ \pi^2 m_*^4}\frac{ \left(2 \text{Re}[\zeta _1 \tilde{\mu }_a]+q_R\right) \left(2 \text{Re}[\zeta _2 \tilde{\mu }_b]+q_R\right)+\left|\zeta _2 \tilde{\mu }_a-\zeta^*
   _1 \tilde{\mu }^*_b+q_S\right|^2{}}{\zeta_1^2+\zeta_2^2+4 } \\
|\gamma_{10, j}|^2=&\,-\frac{c}{m_*^2\pi^2(3-j)!}\frac{1}{\frac{1}{2\text{Re}[\lam_{10}]+q_R(1+|\zeta_{10}|^2)-|m|^2 c^2}+\frac{c^2|\lam_{10}|^2}{|m|^4}\frac{1}{2\text{Re}[\lam_{10}]+q_R(1+|\zeta_{10}|^2)+|m|^2 c^2}}\left (\frac{q_R}{m_*^2}\right )^{4-j}\\
|\gamma_{5, j}|^2=&\,-\frac{c}{m_*^2\pi^2(3-j)!}\frac{1}{\frac{1}{2\text{Re}[\lam_{5}]+q_R(1+|\zeta_{5}|^2)-|m|^2 c^2}+\frac{c^2|\lam_{5}|^2}{|m|^4}\frac{1}{2\text{Re}[\lam_{5}]+q_R(1+|\zeta_{5}|^2)+|m|^2 c^2}}\left (\frac{q_R}{m_*^2}\right )^{4-j}\,.
\end{align}
\end{subequations}
We display the normalisation factors for the model B in appendix \ref{ap:modelB}.

Note that the parameters $\lambda$ and $\zeta$ that appear in the various normalisation factors depend on the local 
flux densities and in particular on the flux hypercharge. This implies that the normalisation factors of the MSSM multiplets sitting
in the same GUT multiplet will be different. As pointed out in \cite{afim11,fimr12,fmrz13} this is  a key feature to obtain realistic mass ratios,
as we will see in section \ref{s:fitting}.

\subsection{Non-perturbative corrections to the wavefunctions}

So far we have been discussing the kinetic terms of the matter fields neglecting non-perturbative corrections. However, as we are computing Yukawa
couplings up to first order in the parameter $\eps$, one should consider the expression for the kinetic terms at the same level of approximation. 
We will discuss now how these effects enter in the computation of the kinetic terms and show that for our models no relevant correction is 
produced. This implies that the result obtained above may be used in the computation of physical Yukawa matrices.

The F-term equations of motion corrected at $\CO (\eps)$ are
\be\begin{split}
\bar \p_{\langle A\rangle} a &=0\,,\\
\bar \p_{\langle A\rangle} \varphi &=  i [a, \langle \Phi\rangle ]- \eps \p \theta_0 \wedge (\p_{\langle A\rangle} a +\bar \p_{\langle A\rangle} a^\dagger)\,,
\end{split}
\ee
which need to be supplemented with the D-term equation \eqref{treeD} which is not affected by non-perturbative corrections \cite{afim11}. In this section we shall
simply show the final result and discuss the impact of non-perturbative corrections on the kinetic terms, deferring the details of the computation to Appendix \ref{ap:holoyuk}.

\subsubsection*{Sectors not affected by T-brane}
In both sectors not affected by the T-brane background the corrections take the same form
\be\label{corr2}\begin{split}
\vec\vphi\, &=\, \gamma\left(\begin{array}{c}-\frac{i \zeta}{2 \tilde \mu_a} \\ \frac{i (\zeta-\lambda)}{2 \tilde \mu_b}\\1\end{array}\right) \, 
e^{\frac{q_R}{2} (x \bar x -y \bar y)-q_S\text{Re} (x \bar y)+ (\mu_a x+ \mu_b y)(\zeta_1 \bar x - \zeta_2 \bar y)}\,\left[f(\zeta_2x+\zeta_1y)+\eps B(\zeta_2 x + \zeta_1 y)+\eps \Upsilon\right]\,.
\end{split}\ee
The function $\Upsilon$ that controls the $\CO(\eps)$ correction is
\be\begin{split}
\Upsilon &= \frac{1}{4} (\zeta_1 \bar x -\zeta_2 \bar y )^2 (\theta_y \mu_a - \theta_x\mu_b) f (\zeta_2 x + \zeta_1 y)+ \frac{1}{2} (\zeta_1 \bar x -
\zeta_2 \bar y)  (\zeta_2 \th_y - \zeta_1 \th_x)f'(\zeta_2 x + \zeta_1 y)+\\
&+ \left[\frac{\delta_1}{2} (\zeta_1 x - \zeta_2 y)^2+\delta_2  (\zeta_1 x - \zeta_2 y) (\zeta_2 x + \zeta_1 y)\right]f(\zeta_2 x + \zeta_1 y)\,,
\end{split}\ee
where
\be\begin{split}
\delta_1 & = \frac{1}{(\zeta_1^2 + \zeta_2^2)^2} \left[\bar \th_x (q_S \zeta_1 -q_R \zeta_2)+ \bar \th_y (q_R \zeta_1 +q_S \zeta_2)\right]\,,\\
\delta_2 & = \frac{1}{(\zeta_1^2 + \zeta_2^2)^2} \left[\bar \th_x (q_R \zeta_1 +q_S \zeta_2)- \bar \th_y (q_S \zeta_1 -q_R \zeta_2)\right]\,.
\end{split}\ee
The holomorphic function $B(\zeta_2 x + \zeta_1 y)$ in \eqref{corr2} is not determined by the equations of motion, and it may be fixed by asking for regularity
of the function $\xi$ that appears in \eqref{eq:Fsol}. We shall nevertheless not  discuss this point here since it does not affect the result for the kinetic terms.

Having the correction it is now possible to discuss the effect of the correction on the kinetic term. We can use the fact that the integrand has to be invariant
under the symmetry $(x,y) \rightarrow e^{i \alpha} (x,y)$ to check whether the corrections actually contribute to the kinetic terms. In the cases when 
the sector hosts a Higgs field (this happens in the model A for both the $\mathbf 5_U$ and the $\mathbf {\bar 5}_D$ and in the model B for the $
\mathbf 5_U$) no correction is generated. In the remaining case, namely the $\mathbf{ \bar 5}_M$  in the model B, there are non-diagonal terms in 
the kinetic terms inducing a mixing between the first and the third families of down quarks and leptons. This however will not affect the computation
of Yukawa couplings, because this $\CO (\eps)$ correction in the kinetic terms will only induce a $\CO(\eps^2)$ correction in the Yukawa matrix. See \cite{fimr12}
for a more detailed discussion of this point in a similar context.

\subsubsection*{Sectors affected by T-brane}
As shown in Appendix \ref{ap:holoyuk} the structure of the solution for both sectors charged under the T-brane background is
\be
\label{phys10np} 
\vec{\vphi}_{{10}^+} \, =\, 
\left(\begin{array}{c} \bullet \\ \bullet \\ 0\end{array}\right)
 + \eps
\left(\begin{array}{c} 0 \\ 0 \\ \bullet\end{array}\right) +\CO(\eps^2)
 \qquad 
\vec{\vphi}_{{10}^-} \, =\,
\left(\begin{array}{c} 0 \\ 0 \\ \bullet\end{array}\right)
 + \eps
 \left(\begin{array}{c} \bullet \\ \bullet \\ 0\end{array}\right)+\CO(\eps^2).
\ee
The peculiar structure of the $\CO (\eps)$ corrections allows us to demonstrate that these corrections will not affect the kinetic terms without needing
to write explicitly their form. Indeed it follows from \eqref{4dnorm} that the corrections vanish because at $\CO (\eps)$ the correction is proportional 
to\footnote{Here the superscript $(0)$ denotes the tree-level term and $(1)$ the $\CO(\eps)$ correction.} $\vec{\vphi}_{{10}^+}^{(0)}\cdot\vec{\vphi}_{{10}^-}^{(1)}$ and $\vec{\vphi}_{{10}^-}^{(0)}\cdot\vec{\vphi}_{{10}^+}^{(1)}$ and these scalar products are both zero. This structure is similar to the one observed in  \cite{fmrz13}, to which we refer the reader for a more detailed discussion of this point.

\section{Fitting fermion masses and mixing angles}
\label{s:fitting}

Gathering the results of the last two sections one may write the final expression for the physical Yukawa matrices at the GUT scale. In particular, for the model A one obtains the matrices (\ref{hyukU}) and (\ref{hyukD}) with the normalisation factors given by \eqref{normA}, while the same quantities for the model B are given in Appendix \ref{ap:modelB}.  As noted above the value of the normalisation factors varies for MSSM field with different hypercharge even if they sit inside the same GUT multiplet, something that we will indicate by adding a superscript to distinguish between them. 

Based on these result in this section we explore  whether it is possible to find some regions in the parameter space of our models where we may reproduce the
realistic values for fermion masses and mixings. Our calculations are performed at the GUT scale which is usually taken around $10^{16}$ GeV and
therefore to compare the values for the fermion masses it is necessary to follow the values of the fermion masses along the renormalisation group flow. 
We show in table \ref{tab:massas} the extrapolation of the fermion masses up to the unification scale taken from \cite{Ross:2007az} in the context of the MSSM. Since in the MSSM
two Higgs fields are present the values depend on an additional parameter $\tan \beta$ which controls how the observed vev of the Higgs is
distributed between the $H_u$ and the $H_d$ Higgs fields of the MSSM. More specifically $\langle H_u\rangle = V \cos \beta$ and $\langle H_d\rangle = V 
\sin \beta$ where $V $ is the measured value of the Higgs field and is given by $V \approx 174 $ GeV. We shall now discuss the comparison between these
extrapolated data and the values for the Yukawa couplings that we obtain in our local $E_7$ models.
\begin{table}[htb] 
\renewcommand{\arraystretch}{1.25}
\begin{center}
\begin{tabular}{|c||c|c|c|}
\hline
tan$\beta$  &  10&   38  &  50 \\
\hline\hline
$m_u/m_c$ &   $2.7\pm 0.6\times 10^{-3}$   &  $2.7\pm 0.6\times 10^{-3}$&$2.7\pm 0.6\times 10^{-3}$  \\
\hline
$m_c/m_t$ &   $2.5\pm 0.2\times 10^{-3}$ &$2.4\pm 0.2\times 10^{-3}$&$2.3\pm 0.2\times 10^{-3}$ \\
\hline\hline
$m_d/m_s$ &   $5.1\pm 0.7\times 10^{-2}$   &  $5.1\pm 0.7\times 10^{-2}$  & $5.1\pm 0.7\times 10^{-2}$  \\
\hline
$m_s/m_b$ &    $1.9\pm 0.2\times 10^{-2}$   &  $1.7\pm 0.2\times 10^{-2}$  & $1.6\pm 0.2\times 10^{-2}$  \\
\hline\hline
$m_e/m_\mu$  &   $4.8\pm 0.2\times 10^{-3}$   &  $4.8\pm 0.2\times 10^{-3}$  & $4.8\pm 0.2\times 10^{-3}$   \\
\hline
$m_\mu/m_\tau$  &    $5.9\pm 0.2\times 10^{-2}$   &  $5.4\pm 0.2\times 10^{-2}$  & $5.0\pm 0.2\times 10^{-2}$  \\
\hline\hline
$Y_\tau $  &    $0.070\pm0.003 $   &  $0.32\pm0.02 $ &    $0.51\pm0.04 $ \\
\hline
$Y_b $  &    $0.051\pm0.002 $   &  $0.23\pm0.01 $ &    $0.37\pm0.02 $ \\
\hline
$Y_t $  &    $0.48\pm0.02 $   &  $0.49\pm0.02 $ &    $0.51\pm0.04 $ \\
\hline
\end{tabular}
\end{center}
\caption{\small Running mass ratios of  quarks and leptons at the unification scale from ref.\cite{Ross:2007az}.}
\label{tab:massas}
\end{table}

\subsection{Fermion masses}

Knowing the Yukawa matrices we can easily extract the values of the fermion masses which depend on the eigenvalues of the matrices. From the Yukawa
matrices in \eqref{hyukU} and \eqref{hyukD} we see that the eigenvalues are
\be\label{Yeigen}
\begin{array}{ll}
Y_t=\g_U\,\g^Q_{10,3}\g^U_{10,3} \,Y_{33}^{U} &  \qquad Y_c=\eps \,\g_U\,\g^Q_{10,2}\g^U_{10,2}\, Y_{22}^U\\
Y_b=\g_D\left(\g^Q_{10,3}\g^D_{5,3} Y_{33}^{D/L}+\eps \, \g_{10,2}^Q \g_{5,2}^D \delta \right) &\qquad  Y_s= \eps\,\g_D \left(\g^Q_{10,2}\g^D_{5,2}Y_{22}^{D/L} - \g_{10,2}^Q \g_{5,2}^D\delta\right) \\
Y_\tau=\g_D\left( \g^E_{10,3}\g^L_{5,3}Y_{33}^{D/L}+\eps\,\g_{10,2}^E \g_{5,2}^L  \delta \right) &\qquad  Y_\mu=\eps\, \g_D\left(\g^E_{10,2}\g^L_{5,2} Y_{22}^{D/L}-\g_{10,2}^E \g_{5,2}^L  \delta \right)
\end{array}
\ee
while for the first family we have that
\be
Y_u, \ Y_d, \ Y_e \, \sim \, \mathcal O(\eps^2)
\ee
Here the normalisation factors are those given in the previous section, and we have defined
\be
\delta = - \tilde \kappa\,\frac{\pi ^2 d    (a-b) \left[\theta _y (a (d-2)-b (4 d+1))-3
   (d+1) \theta _x\right]}{(d+1)^5  \rho _{\mu } \,\rho _m^{5/2}}
\ee
Therefore we see that when $a \neq b$ the eigenvalues of the down-type Yukawa matrix are different from the diagonal entries of the matrix. However we note
that this correction will be of order $\mathcal O (\tilde \kappa)$, which at the end of this section will be fixed to be $10^{-5}-10^{-6}$ by fixing the value of the quark mixing angles. In this sense we can neglect $\d$ as compared to the contribution coming from the diagonal entries of the down-type Yukawa matrix, as well as any $\tilde \kappa$ dependence on these entries. After this it is easy to see manifestly the $(\CO (1),\CO (\eps), \CO(\eps^2))$ hierarchy between the three families of quarks and leptons. Because the explicit expression for the eigenvalues of the lightest family cannot be computed at the level of approximation that we are working, we turn to discuss the masses for the two heavier families.

\subsubsection*{Masses for the second family}

The strategy that we choose to follow to see if it is possible to fit all fermions masses is to look first at the mass ratios between the second and third 
families, which do not depend on $\tan \beta$. More specifically we will start by considering the following mass ratios
\be\label{eq:ratior}
\frac{m_\mu/m_\tau}{m_s/m_b}\,, \qquad \frac{m_c/m_t}{m_s/m_b}\,,
\ee
which, in addition to being independent of $\tan \beta$ do not depend on the parameter $\eps$ which measures the strength of the non-perturbative
effects. From the data in table \ref{tab:massas} and the discussion in \cite{fimr12,fmrz13} we aim to reproduce the following values
\be
\frac{m_\mu/m_\tau}{m_s/m_b} = 3.3\pm 1 \,, \qquad \frac{m_c/m_t}{m_s/m_b} = 0.13\pm 0.03\,.
\label{rvalues}
\ee

To complete the discussion of the masses of the second family we can look at an additional mass ratio, namely $m_c/m_t$. Being able to correctly fix this  
quantity and (\ref{rvalues}) allows us to obtain correct mass values for the second family of quarks and leptons when the masses of the third family are fitted later on.

Let us now discuss the behaviour of these particular ratios of masses in the two models we have been discussing so far. We will see already at this 
level that the model B does not allow for good values of these ratio of masses. 

\paragraph{Model A}

We can compute the aforementioned ratios for the case of the model A and the result is
\begin{subequations}\label{ratioss}
\begin{align}
\frac{m_c}{m_t}\,=&\,\,\left| \frac{\tilde\eps}{2\rho_\mu}\right|  \sqrt{ q_{R}^Q\,q_{R}^{U} }\,=\,  \left|\frac{\tilde\eps \,\tilde N_Y}{2 \rho_\mu}\right| \sqrt{\left (x-\frac{1}{6}\right )\left (x + \frac{2}{3}\right )} \\
\frac{m_s}{m_b}\,=&\,\, \frac{Y_s}{Y_b}\,   \sqrt{ q_{R}^{Q}\,q_{R}^{D} }\,\simeq\, \frac{\left| \tilde{N}_Y\right|  \left[(d
+1)\theta _x+(a+b d )\theta _y\right]}{(d+1)^2 \rho _{\mu }}\sqrt{\left( x-\frac{1}{6} \right) \left(  y-\frac{1}{3}\right)}
  \\
\frac{m_\mu}{m_\tau}\,=&\,\, \frac{Y_\mu}{Y_\tau}\,   \sqrt{ q_{R}^{Q}\,q_{R}^{D} }\,\simeq\, \frac{\left| \tilde{N}_Y\right|  \left[(d
+1)\theta _x+(a+b d )\theta _y\right]}{(d+1)^2 \rho _{\mu }} \sqrt{\left (x-1\right )\left (y +\frac{1}{2}\right )}
\end{align}
\end{subequations}
where we defined \be x = -\frac{M_1}{\tilde N_Y}\,, \quad y = -\frac{M_2}{\tilde N_Y}\,, \quad d = \frac{\mu_2^2}{\mu_1^2}\,.\ee
In writing the final expression for $m_s/m_b$ and $m_\mu/m_\tau$ we neglected the $\delta$ shifts appearing in the expressions for the 
eigenvalues of the down quark and lepton Yukawa matrix as well as the $\CO(\eps)$ correction appearing in $Y_{33}^{D/L}$. The reason behind this choice is that these contributions are 
much smaller when compared to the other terms and therefore will not affect the final results. Once these contributions are neglected the expressions
for the ratio of masses become much simpler and depend on a smaller subset of parameters giving therefore more analytical control.
Using \eqref{ratioss} we can compute the ratio of masses \eqref{eq:ratior} and the results are
\be\label{eq:ratio1}
\frac{m_\mu/m_\tau}{m_s/m_b} = \sqrt{\frac{(x-1)\left(y-\frac{1}{2}\right)}{\left(x-\frac{1}{6}\right)\left(y-\frac{1}{3}\right)}}\,,
\ee
\be\label{eq:ratio2}
\frac{m_c/m_t}{m_s/m_b} = \frac{(d+1)^2 \sqrt{2 +3 x} (a \theta_y+\theta_x)}{2 \sqrt{3 y-1}  \left[(d
+1)\theta _x+(a+b d )\theta _y\right]}\,,
\ee

\begin{figure}[ht]
\center\includegraphics[height=7cm]{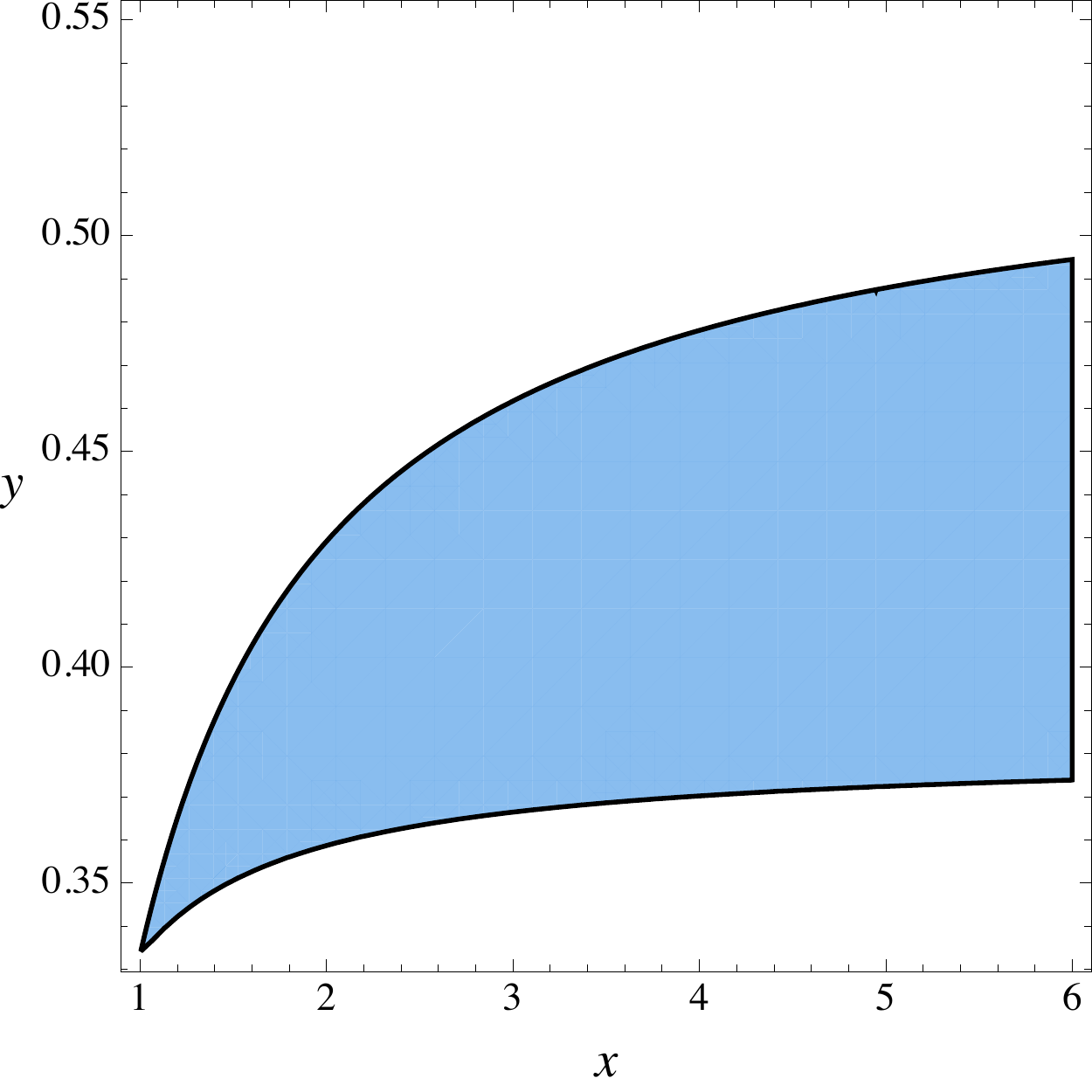}  \includegraphics[height=7cm]{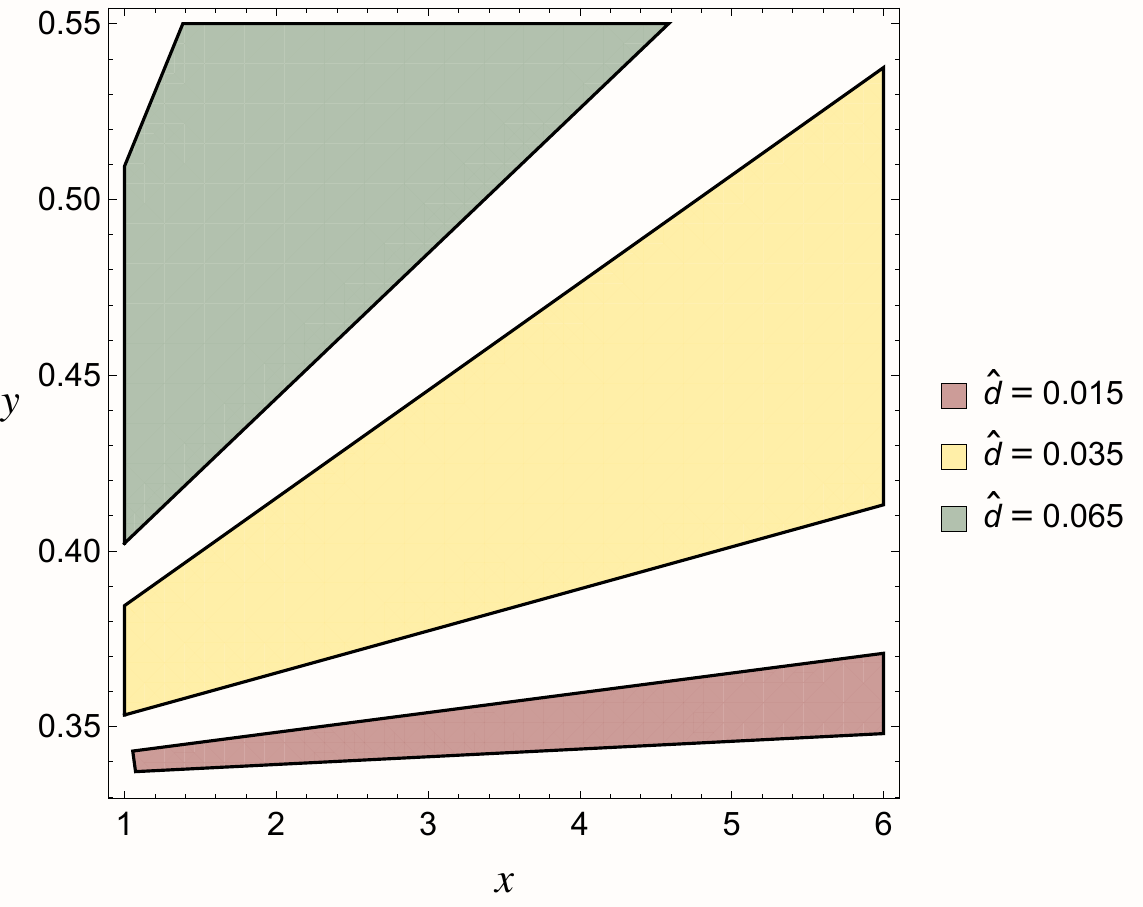}
 \caption{\small{On the left the region in the $x-y$ plane for the ratio of masses \eqref{eq:ratio1} compatible with the realistic value in (\ref{rvalues}). On the
 right the region in the $x-y$ plane for the ratio of masses \eqref{eq:ratio2} compatible with (\ref{rvalues}), for different values of $\hat d$.}}
 \label{fig:ratiosa}
\end{figure}

\begin{figure}[ht!]
\center\includegraphics[height=6.5cm]{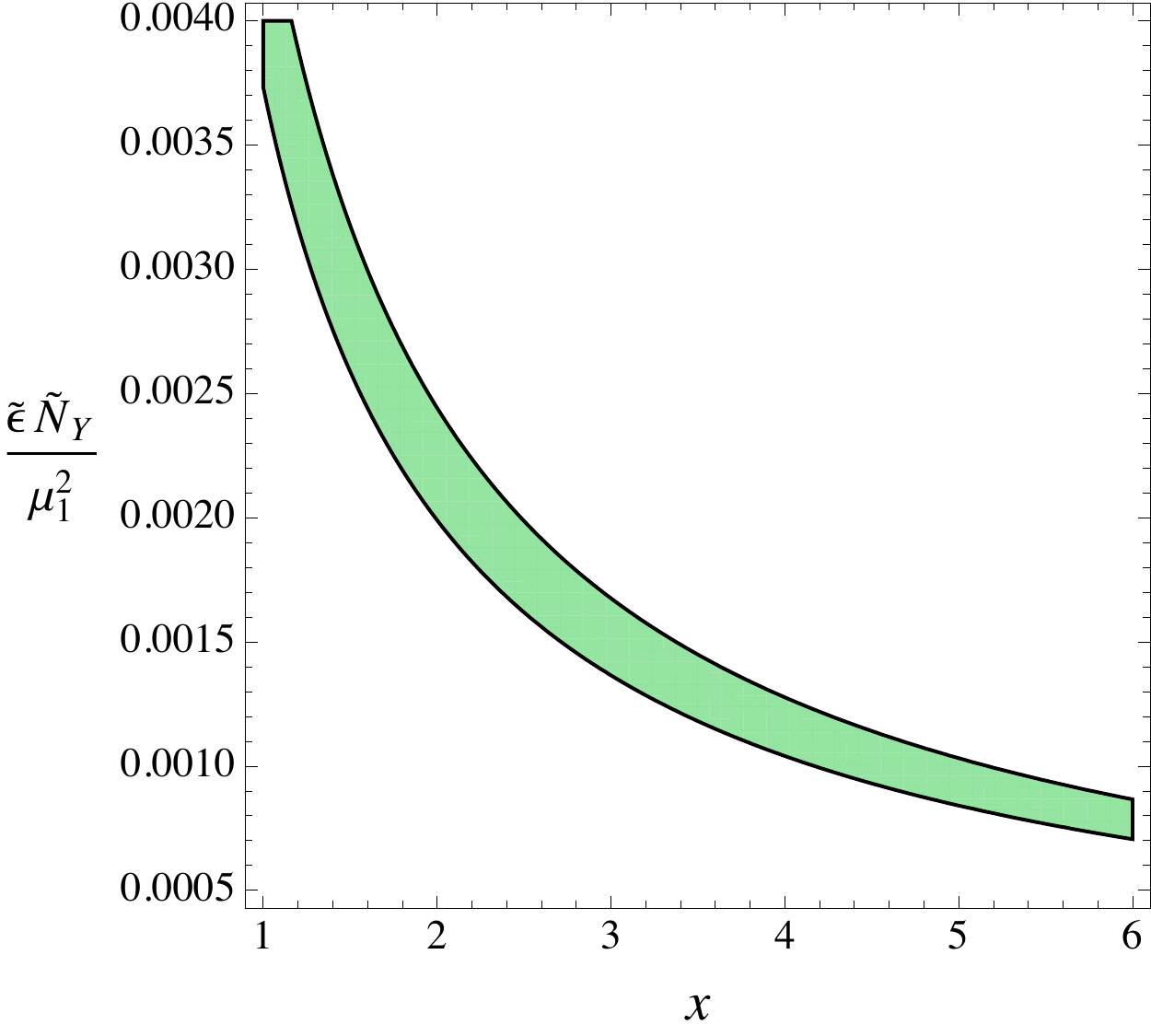}
 \caption{\small{Region in the plane $x-\tilde \eps \tilde N_Y/\mu_1^2$ for the ratio \eqref{eq:ratio3} to be compatible with the 
range of values in table \ref{tab:massas}.}}
 \label{fig:ratiosb}
\end{figure}

Chirality conditions place some constraints in the allowed regions for $x$ and $y$, and in particular we find that for $\tilde N_Y<0$ we need
$x<-2/3 $ and $y<1/2$  and for $\tilde N_Y>0 $ we need $x>1$ and $y>1/3$. Between the two possibilities we find that it is simpler to fit the empirical 
data by choosing $\tilde N_Y >0$. Moreover it seems reasonable to take $\th_x \sim \th_y$ which implies
that both ratios of masses will depend only on three parameters, namely $x$, $y$ and $\hat d$ where
\be
\hat d = \frac{(d+1)^2(a+1)}{a+1+d(b+1)}\,.
\ee
 We show in figure  \ref{fig:ratiosa} of the $x$ and $y$
parameter space where we find values for the ratios of masses in agreement with the empirical ones.
The remaining mass ratio $m_c/m_t$ has also a 
nice analytical expression in terms of the parameters of our local model
\be\label{eq:ratio3}
\frac{m_c}{m_t} = \frac{\sqrt{\left(x-\frac{1}{6}\right) \left(\frac{2}{3}+x\right)} | \tilde N_Y|}{2
   \mu_1^2}\, \tilde \eps\,.
\ee
In figure \ref{fig:ratiosb} we show in which region of the $x$ and $\tilde \eps\,| \tilde N_Y|/ \mu_1^2$ parameter space we are able to find good
values for this last ratio of masses.

\begin{figure}[ht]
\center\includegraphics[width=9cm]{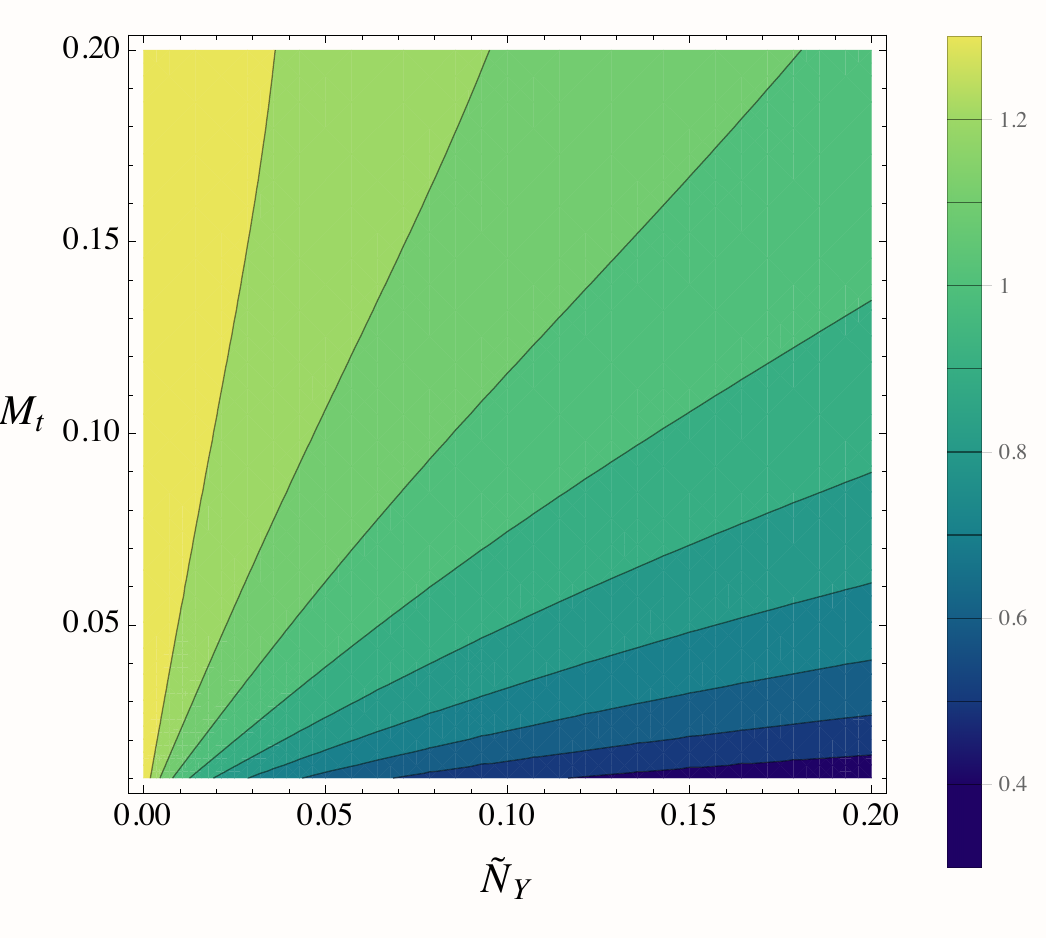}
 \caption{\small{Value of the ratio of ratios $(m_\mu/m_\tau)/(m_s/m_b)$ in the model B in the $\tilde N_Y - M_t$ plane, where are taking $M_t =- M_1 - \tilde N_Y > 0$ and $\tilde N_Y >0$ as dictated by eq.(\ref{condBmodel}).}}
 \label{fig:B}
\end{figure}

\paragraph{Model B} Contrary to model A, for the model B one does not find simplified expressions for the fermion mass ratios. We have explored numerically different regions in parameter space trying to reproduce the value in (\ref{rvalues}) for the ratio of ratios $(m_\mu/m_\tau)/(m_s/m_b)$ without success. In fact, in figure \ref{fig:B} shows how trying to achieve a realistic value for this pushes us to a region of the parameter space in which $\tilde N_Y$ is negative, which is in conflict with the condition (\ref{condBmodel}) necessary for a realistic chiral spectrum in this model. It would be interesting to have a more intuitive understanding of why this model fails to reproduce the empiric data as compared to case of the model A.

\subsubsection*{Yukawa couplings for the third family}

\begin{figure}[ht]
\center\includegraphics[height=7cm]{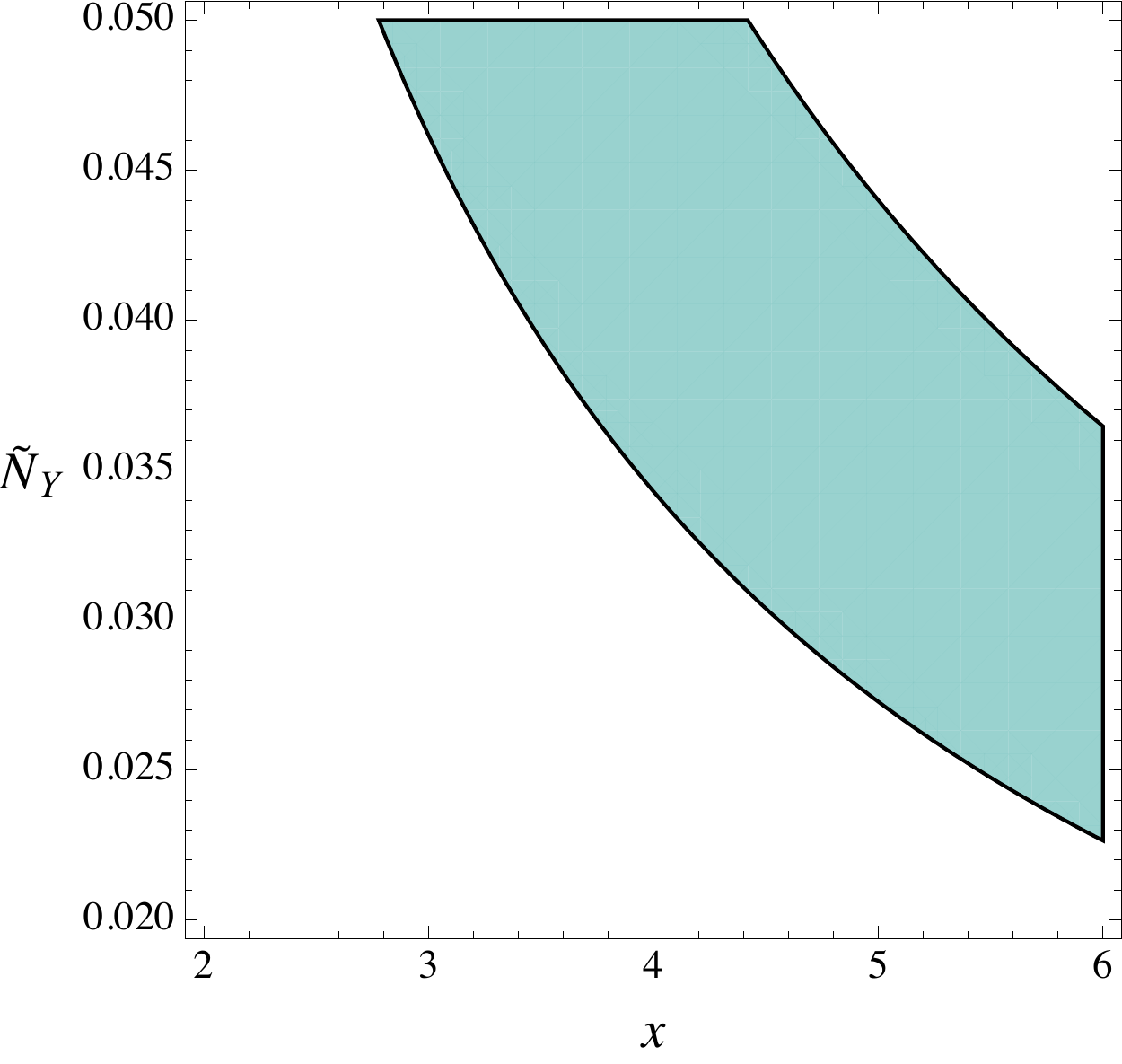}
 \caption{\small{Region in the $x-\tilde N_Y$ plane with a ratio $Y_\tau/Y_b$ compatible with table \ref{tab:massas}.}}
 \label{fig:ratiotaub}
\end{figure}

Given that in the case of the model A we have been able to find regions where the mass ratios between the second and third families are 
compatible with the MSSM, all we need to fix now are the masses for the fermions in the third family. We start by looking at the ratio between
the mass of the $\tau$-lepton and the b-quark. Such ratio can be expressed in terms of normalisation factors only
\be
\frac{Y_{\tau}}{Y_b} = \frac{\g_{10,3}^E \g_{5,3}^L}{\g_{10,3}^Q \g_{5,3}^D}\,,
\ee
but in terms of the model parameters it acquires a rather complicated form, so it is quite hard to describe analytically the region of parameter space that is compatible with the expected value
\be
\frac{Y_\tau}{Y_b} = 1.37\pm 0.1 \pm0.2\,.
\ee
We have therefore performed a numerical scan over the values of the local flux densities which are compatible with the conditions for chirality and doublet-triplet splitting, and with the fermion mass ratios just discussed. More precisely we have chosen the following  point in parameter space\footnote{We normalise all local flux densities in units of $m_{st}^2$ where the string scale $m_{st}$ is related to the typical F-theory scale $m_*$
by $m_{st}^4 = (2\pi)^3 g_s m_*^4$. In all the computations done in this section we take $g_s \sim \CO(1)$.}
\be\label{pointp}\begin{split}
(\rho_m,\rho_\mu,d,c,a,b, \eps \,\th_x,\eps \,\th_y) &=(0.23, 2.5\times 10^{-3}, -0.9, 0.25,-0.4,-0.6,  10^{-4},10^{-4})\\
(M_1,M_2,N_1,N_2,\tilde N_Y,N_Y)& = (-0.17, -0.0136, -0.14, 0.008, 0.034, 0.1953)
\end{split}\ee
and scanned over the allowed values for $x$ and $\tilde N_Y$ that do not spoil the constraints above. We show our results in figure \ref{fig:ratiotaub}, which displays a rather large region of these parameters.

\begin{figure}[htb!]
\center\includegraphics[width=10cm]{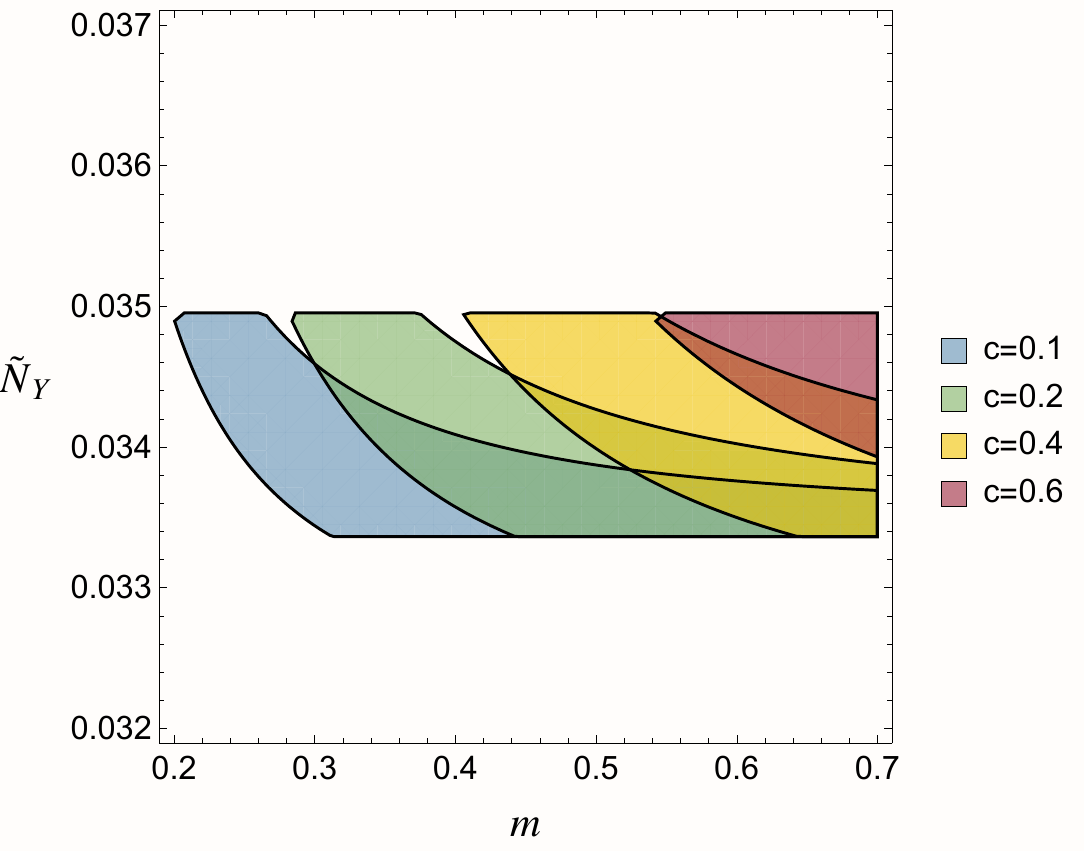}
 \caption{\small{Regions in the $m$-$\tilde N_Y$ plane where all constraints are fulfilled for different values of $c$.}}
 \label{fig:total}
\end{figure}

Finally we may wish to see whether all constraints for chirality, doublet-triplet splitting and realistic fermion mass ratios may be solved simultaneously. We find that this is true for large regions of the parameter space. To illustrate this fact, in figure \ref{fig:total} we plot regions in the $m-\tilde N_Y$ parameter space where 
all constraints are fulfilled for different values of $c$. By inspecting the plot we see that regions fulfilling all constraints exist for different values of $c$
which are of the same order as \eqref{eq:cvalue}.
\begin{figure}[htb!]
\center\includegraphics[width=10cm]{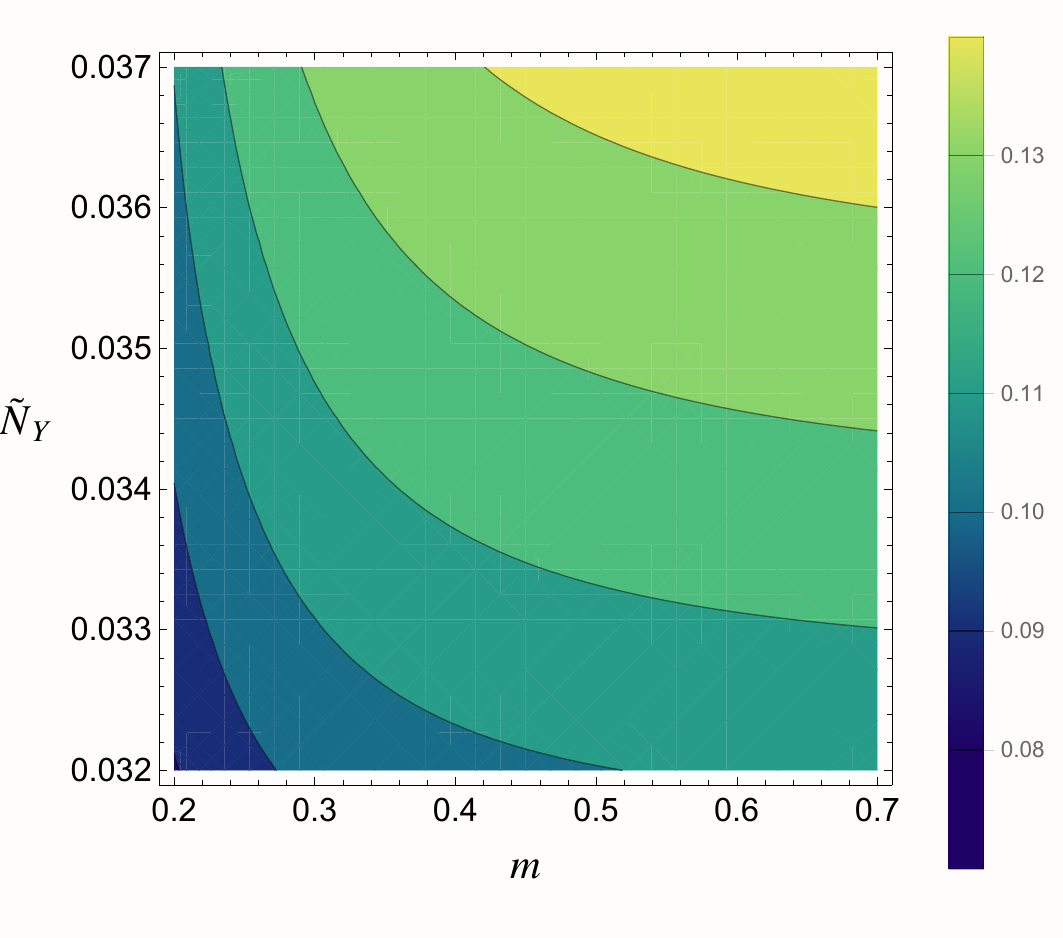}
 \caption{\small{Value of $Y_b$ in the $m-\tilde N_Y$ plane with the other parameters fixed.}}
 \label{fig:beauty}
\end{figure}

In these regions we can look at the typical value of the b-quark Yukawa to estimate the value of $\tan \beta$ that we
typically obtain from our scan. We show in figure \ref{fig:beauty} the possible values of $Y_b$
and by comparison with the content of the table \ref{tab:massas} we obtain an approximated value of $\tan \beta \simeq 10-20$ .

\subsubsection*{Comparison with previous scans}

While the Yukawa couplings just discussed arise from the $E_7$ model built in section \ref{s:e7model}, they are in fact more general, in the sense that they also correspond to certain models with $E_8$ enhancement. In particular, as mentioned below eq.(\ref{sigmas}) the matter curve content containing the MSSM chiral fields is identical to the one of the $E_8$ model explored in \cite{mrz15}.\footnote{In such model the T-brane structure of $\langle \Phi \rangle$ is more complicated, but however the matter sectors containing the MSSM chiral content are unaffected by such extra structure. Therefore one can directly apply the computation of Yukawa couplings  performed in this paper to such local $E_8$ model.} More precisely, we have that we recover the matter curves and the Yukawas of such $E_8$ model if in the parameters that describe the Higgs vev in (\ref{Phixy}) we fix $a=b=1$. As mentioned in section \ref{sec:flux}, such particular choice of parameters prevents to implement the doublet-triplet splitting mechanism that by means of the hypercharge flux threading the matter curves {\bf 5}$_U$ and {\bf 5}$_D$ \cite{bhv2}. 

\begin{figure}[ht]
\center\includegraphics[width=13cm]{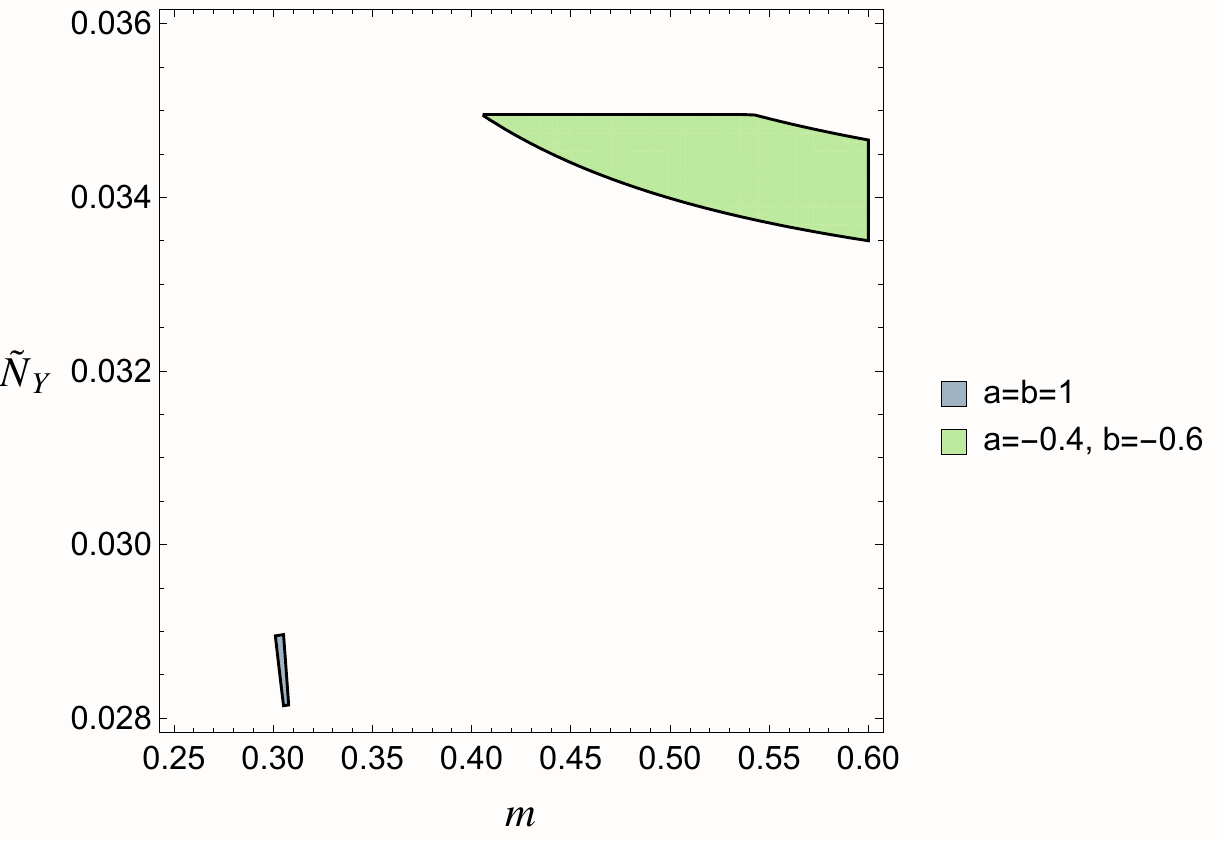}
 \caption{\small{Comparison between the current scan and the one in \cite{mrz15}, which considers $a=b=1$.}}
 \label{fig:E7E8}
\end{figure}
We find quite intriguing that, by opening this new directions in parameter spaces that allow for doublet-triplet splitting, we are also able to fit the fermion masses much more easily than in previous attempts. We have illustrated such effect by means of figure \ref{fig:E7E8}, where we plot the allowed regions for realistic fermion masses in the case of the current scan and the one performed for the model in \cite{mrz15}, which assumes $a=b=1$. For illustrative purposes we have again used the plane $\tilde N_Y - m$ of parameters, but the fact that the region where realistic fermion masses are reproduced is wider in this case than the one in \cite{mrz15} is true for any direction in parameter space. Finally, a further advantage of exploring this new region of parameters is that the worldvolume flux densities are now much lower than in previous cases (c.f. eq.(\ref{pointp}) as compared to eq.(6.12) in \cite{mrz15}). 
Being in the regime of diluted fluxes is quite important to construct 7-brane local models where $\alpha'$ corrections are negligible, and therefore the 7-brane action of \cite{bhv1,bhv2} can be used reliably. In particular, for models where the flux densities are larger than $m_*$ one may worry that the D-term (\ref{FI7}) receives non-trivial corrections that could modify the computation of wavefunctions in the real gauge. 

\subsection{Quark mixing angles}

An additional piece of information that we may extract from the Yukawa matrices involves the quark mixing angles, which are conventionally encoded
in the CKM matrix. The definition of the CKM matrix involves a pair of unitary matrices $V_U$ and $V_D$ which diagonalise the product $Y Y^\dag$
of the quark Yukawa matrices. More specifically we have that 
\begin{subequations}\label{uves}
\begin{align}
M_U=&\,V_U Y_UY_U^\dagger V_U^\dagger\\
M_D=&\,V_D Y_DY_D^\dagger V_D^\dagger
\end{align}
\end{subequations}
with $M_U$ and $M_D$ diagonal. Using this we may define the CKM matrix as
\be
V_{CKM}=V_UV_D^\dagger\,.
\ee
We can directly apply this definition to the Yukawa matrices of our model, which are accurate up to $\CO (\eps^2)$ corrections. In general the result is quite complicated, but again we find that we recover the CKM structure of \cite{mrz15} when we set $a=b$. To compare to the results therein we expand our more general CKM matrix in the new parameter $\xi \equiv a-b$. From the above analysis we know that for realistic fermion mass values $|\xi| \sim 0.1$, and so this expansion will quickly converge. 

Explicitly we find 
\begin{subequations}
\begin{align}
\hat V_U=&\,\left (\begin{array}{ccc}
1&0&-\frac{\tilde\eps\g^Q_{10,1}}{2\rho_\mu\g^Q_{10,3}}\\
0&1&0\\
\frac{\tilde\eps^*\g^Q_{10,1}}{2\rho^*_\mu\g^Q_{10,3}}&0&1 \end{array}\right )\\
\hat V_D =& \  \ \hat V_D^{(0)} + \xi \hat V_D^{(1)} + \CO(\xi^2)\\
\hat V_D^{(0)} =&\left(
\begin{array}{ccc}
 1 &%
 -\,\frac{   i \tilde \epsilon\text{Im} \left[(d+1) \tilde \kappa^* \rho_\mu\right]}{ (d+1)|d+1|^2
  |\rho _{\mu }|^2 \rho _{\mu }}\frac{\g_{10,1}^Q \g_{10,2}^Q}{(\g_{10,3}^Q)^2} &%
   \frac{(d+1) \tilde \epsilon   \rho _{\mu }-2 \tilde \kappa ^2}{ (d+1)^2 \rho
   _{\mu }^2} \frac{\g_{10,1}^Q}{\g_{10,3}^Q}\\%
 -\frac{\tilde{\epsilon }^* \tilde{\kappa }^*}{ \left(d^*+1\right)^2
{\rho^{*2}_{\mu }}{}} \frac{\g_{10,1}^Q\g_{10,2}^D}{(\g_{10,3}^Q)^2}&%
    1-\frac{|\tilde \kappa|^2}{2 |d+1|^2
| \rho_\mu|^2} \frac{(\g_{10,2}^Q)^2}{(\g_{10,3}^Q)^2} &%
   \frac{\tilde \kappa }{(d+1) \rho _{\mu }}\frac{\g_{10,2}^Q}{\g_{10,3}^Q} \\%
 -\frac{\left(d^*+1\right) \tilde \eps^*
{\rho^* _{\mu }}{} -2\tilde \kappa^{* 2}}{2
   \left(d^*+1\right)^2 {\rho^* _{\mu }}^2}\frac{\g_{10,1}^Q}{\g_{10,3}^D} &%
   -\frac{\tilde \kappa ^*}{\left(d^*+1\right)
{\rho^* _{\mu }}} \frac{\g_{10,2}^Q}{\g_{10,3}^Q}&%
    1-\frac{|\tilde \kappa|^2}{2 |d+1|^2
| \rho_\mu|^2} \frac{(\g_{10,2}^Q)^2}{(\g_{10,3}^Q)^2}\\
\end{array}
\right)\\
\hat V_D^{(1)}=&\left(
\begin{array}{ccc}
 0 &%
  \frac{(d-1) \epsilon  \tilde{\kappa }^* \theta _y}{2
   |d+1|^2 \left({d}+1\right) |\rho _{\mu }|^2
   {}} \frac{\g_{10,1}^Q \g_{10,2}^Q}{(\g_{10,3}^Q)^2}&%
    \frac{ 
   20 d^2 \tilde \kappa 
   \rho _{\mu }\tilde \eps-\eps\theta _y \left(d^2-1\right) \rho _m^{3/2}}{2 (d+1)^3 \rho _{\mu
   } \rho _m^{3/2}}\frac{\g_{10,1}^Q}{\g_{10,3}^Q} \\
 0 &%
  \frac{3 d^2 \tilde\epsilon  \tilde{\kappa }^* \rho _{\mu }
   }{ (d+1)
   |{d}+1|^2 \rho _m^{3/2} {\rho^{*} _{\mu
   }}{}} \frac{(\g_{10,2}^Q)^2}{(\g_{10,3}^Q)^2} & %
   -\frac{d \left(3 d (d+1) \tilde \epsilon  \rho _{\mu
   }+4 \kappa ^2\right)}{ (d+1)^3 \rho _m^{3/2}} \frac{\g_{10,2}^Q}{\g_{10,3}^Q} \\%
 -\frac{20{d}^{*2} 
   \tilde \kappa^*\tilde \eps^*{\rho^* _{\mu }}{} +\eps\,\theta _y {\rho^{*3/2}
   _m(1-d^{*2})}{}}{2 \left({d}^*+1\right)^3
{\rho^{*} _{\mu }}{\,} {\rho^{*3/2}_m}{}} \frac{\g_{10,1}^Q}{\g_{10,3}^D}&%
   \frac{d^* \left(3 d^* \left(d^*+1\right)
   \tilde{\epsilon}^* {\rho^* _{\mu }}{} +4 \bar{\kappa
   }^2\right)}{ \left({d}^*+1\right)^3 {\rho^{*3/2}
   _m}{}} \frac{\g_{10,2}^Q}{\g_{10,3}^Q}& %
   \frac{3 \tilde \kappa  d^{*2} \tilde{\epsilon }^*
   {\rho^{*} _{\mu }}{}}{ |d+1|^2 \left(\bar{d}+1\right) \rho
   _{\mu }\rho _m^{*3/2}}\frac{(\g_{10,2}^Q)^2}{(\g_{10,3}^Q)^2}  \\
\end{array}
\right)
\end{align}
\end{subequations}
and one can check that $\hat V_D^{(0)}$ is similar to the rotation matrix found in \cite{mrz15}, with a strong dependence on the parameter $\tilde \kappa$. As in there, one can estimate the effect of $\CO(\eps^2)$ corrections to the Yukawa matrices and kinetic terms by means of some unknown rotation matrices of the form 
\be
V_U = R_U \hat V_U\,, \qquad V_D = R_D \hat V_D\,,
\ee
where
\be
R_{U,D} = \left(\begin{array}{c c c}1 & \eps^2 \alpha_{U,D} &0\\ -\eps^2\alpha_{U,D}&1&0\\0&0&1 \end{array}\right)\,,
\ee
and $\alpha_{U,D}$ are some unknown coefficients. The effect of these rotations is to modify the elements of the CKM matrix involving the first family of quarks but will not affect the mixing between the top and bottom quarks. Such mixing can be measured in terms of the $V_{tb}$ entry of the CKM matrix which at the level of approximation we are working is given by
\be
V_{tb} =  1-\frac{|\tilde \kappa|^2q_{R,Q} }{2 |d+1|^2 | \rho_\mu|^2 } 
\left[1+ \,\xi\, \frac{6  \tilde{\epsilon }^* d^{*2} 
   {\rho^{*2} _{\mu }}{}}{  \left(\bar{d}+1\right) \rho _m^{*3/2}\tilde \kappa^*}\right]\,,
\ee
where the factor multiplying $\xi$ takes the value $4\times 10^{-3}$ when we substitute the typical values (\ref{pointp}), and so it corresponds to a negligible correction to the case $\xi = 0$ considered in \cite{mrz15}. The experimental value for this entry of the CKM matrix is 
\be
|V_{tb}|_{exp} \simeq 0.9991
\ee
and so it can be reproduced by taking $\tilde \kappa \sim 10^{-5} - 10^{-6}$. Notice that this value is quite different from the one obtained in \cite{mrz15}, but this mismatch is only due to the different parametrisation of the distance between the Yukawa points $p_{\rm up}$ and $p_{\rm down}$ taken in that reference and in the present one (c.f. footnote \ref{dictionary}). The physical quantity is the distance between these two points in units of the typical scale of $S_{\rm GUT}$, which we can estimate by looking at the $V_{tb}$ entry of the CKM matrix. In fact we have the following relation\footnote{Here we again discard the $\CO (\xi)$ term in the expression for $V_{tb}$ as it is negligible.}
\be
\sqrt{1-|V_{tb}|}\, \simeq \frac{|\tilde \kappa| \sqrt{q_{R,Q}}}{\sqrt{2}|\rho_\mu||d+1|}\,\propto\, m_* \left| \frac{a+bd}{d+1} x_0 - y_0\right|
\ee
where $(x_0 ,y_0)$ are the coordinates of the down Yukawa point, see \eqref{pdown}. This implies that the separation of the two points directly controls
the mixing between the second and third family. In the case $a=b$ (where we recover the CKM matrix of \cite{mrz15}) this separation is measured  along the coordinate $a x-y$ which is precisely the complex coordinate entering in the matter wavefunctions, see \eqref{haches}. In this case the whole effect of mixing is due to a mismatch in the wavefunctions bases between the two points as in \cite{Randall:2009dw}. It would be however interesting to have an intuitive picture for the general case $a \neq b$. In any case, as in \cite{mrz15} using the relation between the measured $V_{tb}$ entry of the CKM matrix and the relative distance between the two Yukawa points we can directly estimate the latter and see that it is of the order of $10^{-2} V_{GUT}^{1/4}$. Hence we can see explicitly that the distance between the two points is rather small when compared to the typical size of $S_{GUT}$ as claimed in \cite{hv08}.

\section{Conclusions}
\label{s:conclu}

In this paper we have analysed the structure of the Yukawa matrices of quarks and leptons in the context of F-theory $SU(5)$ GUTs.
The generation of all the Yukawa matrices in the same local patch of $S_{GUT}$ requires a local enhancement of the gauge group to 
either $E_7$ or $E_8$ and we considered the possible models that may be embedded in the former. We have seen that
among the set of possible models only one shows a promising structure for the Yukawa matrices. Since these Yukawas are essentially
the same ones as found in the context of local $E_8$ models in \cite{mrz15}, our results point towards some sort of universal structure 
for realistic Yukawas  in the context of the proposal made in \cite{mm09}. All these models require the presence of a non-commuting Higgs background
to generate a large mass for the the third family of quarks and leptons and the deformation of the 7-brane superpotential due to non-perturbative effects to
generate a mass for the first two families, creating a flavour hierarchy in agreement with experimental measurements.

The details concerning the 7-brane background and fluxes for the local model were discussed in section \ref{s:e7model}. 
In particular, a sufficiently rich set of fluxes was considered in order to obtain a realistic local 4d chiral spectrum, break the $SU(5)$ gauge group down to $SU(3)\times SU(2) \times U(1)_Y$ and implement double-triplet splitting as in \cite{bhv2}. The last feature is an improvement with a previous attempt to 
obtain a realistic spectrum out of a local patch, and involves considering a more general background compared to the one in \cite{mrz15}. 
Following similar techniques as those applied in \cite{afim11,fimr12,fmrz13,mrz15} we have computed the holomorphic Yukawa 
couplings of this model, taking into account non-perturbative effects, and shown that they exhibit an appropriate hierarchical structure. 
Finally, we were able to obtain the physical Yukawas by means of computing the kinetic terms for the MSSM chiral fields and
imposing their canonical normalisation.

These technical results have led to the analysis of section \ref{s:fitting}, where the phenomenological possibilities for two slightly 
different models with the same structure for the matter curves are studied in detail. In one of the models we find wide regions in 
the parameter space where the values of the Yukawa couplings are compatible with measured values (model A), while in the other we 
do not find any compatible region (model B).\footnote{One may actually interpret this result in terms of the models of $E_8$ enhancement discussed in \cite{mrz15}. In such framework one finds that the Higgs background yielding realistic fermion masses corresponds to a reconstructible T-brane structure which in the fundamental of $SU(5)_\perp$ has the split $2+2+1$, and with a curve assignment similar to model A. Interestingly, this setup is precisely the one highlighted by the analysis in \cite{mrz15}. The curve assignment corresponding to model B was there discarded due to its difficulties to yield realistic masses for the neutrino sector. We find quite remarkable that the results of the present work, fully insensitive to the neutrino sector, further support the choice made in \cite{mrz15}.} As in \cite{mrz15} we also find that the empirical value for the mixing $|V_{tb}|$ implies very
small values for the up and down-type Yukawa points, justifying the initial hypothesis of $E_7$ enhancement. The other entries 
of the CKM matrix are more difficult to analyse, as they heavily depend on non-holomorphic data related to the lightest family
of quark and leptons over which we have poor control in this ultra-local approach. In particular, as argued in \cite{mrz15} one 
would expect that curvature effects within $S_{\rm GUT}$ could play an important  r\^ole in their evaluation, which following the 
general arguments in \cite{hv08} could give a rationale for the size of the Cabibbo angle. 

 The results obtained here show how $SU(5)$ F-theory GUTs possess an interesting and potentially viable flavour structure
 when the proposal of \cite{mm09} is implemented in realistic local models.
 To reach a more precise understanding of this flavour structure it would be necessary to go beyond the leading
 $\eps$ contributions considered so far. This would allow to compute the mass of the first generation of fermions as well as
 additional entries in the CKM matrix. Moreover, in addition to the non-perturbative effects that have been considered in this paper, other effects
 recently studied in \cite{Martucci:2015dxa} may have an impact in the structure of the Yukawa matrices. It would be therefore desirable to develop in more detail the
 computation of the couplings generated by such effects and see if, whenever present, they are comparable, dominant or subdominant with respect to the ones considered here. Finally it would be important to see whether they are can give rise to novel features in the flavour sector of F-theory GUTs.
  Another missing ingredient in this construction is the realisation of the local models considered in this paper in a fully-fledged F-theory compactification. Extension to global models would be necessary to check the possibility of having the correct chiral spectrum in 4d and also the viability of hypercharge flux GUT breaking. Moreover, it may be that parameters that look independent from a local viewpoint are not such globally, something that is crucial for interpreting our results in the context of the landscape of F-theory vacua.
  
 Finally, while we have gained a good insight over the structure
 of couplings of quarks and charged leptons, neutrinos and the MSSM Higgs fields remain elusive. A natural mechanism to generate 
 mass terms for them would be via the coupling to singlets which would eventually get a vev. The presence of singlets which
 are not localised on the GUT divisor makes the computation more involved as the methods employed so far would not be 
 sufficient. It would be therefore desirable to develop techniques for these kind of computations as these missing terms play
 an important r\^ole for flavour physics and electroweak symmetry breaking. We hope to return on these points in the 
 future.

\bigskip

\bigskip

\centerline{\bf \large Acknowledgments}

\bigskip

We would like to thank  D. Regalado for useful discussions. 
This work has been partially supported by the grant FPA2012-32828 from the MINECO, the REA grant agreement PCIG10-GA-2011-304023 from the People Programme of FP7 (Marie Curie Action), the ERC Advanced Grant SPLE under contract ERC-2012-ADG-20120216-320421 and the grant SEV-2012-0249 of the ``Centro de Excelencia Severo Ochoa" Programme. F.C. is supported through a fellowship of the international programme ``La Caixa-Severo Ochoa". F.M. is supported by the Ram\'on y Cajal programme through the grant RYC-2009-05096. G.Z. is supported through a grant from ``Campus Excelencia Internacional UAM+CSIC".
F.M. would like to thank HKUST IAS, the CERN TH Division, the Aspen Center for Theoretical Physics (supported through the NSF grant PHY-1066293) and UW-Madison for hospitality and support during the completion of this work. F.C. and G.Z. would like to thank GGI in Florence for hospitality
during the completion of this work.

\clearpage

\appendix


\section{$E_7$ machinery}
\label{ap:e7}

The Lie algebra of $E_7$ has $133$ generators $Q_{\alpha}$. We will always work in the Weyl-Cartan basis, where such generators are split in the $7$ generators of the Cartan subalgebra $H_i, i=1...7$ and $126$ roots $E_{\rho}$. In this basis he commutation rules among Cartan and roots are the following
\begin{align}
[H_i,E_{\rho}]=\rho_iE_{\rho} \label{rootdef}
\end{align}
From (\ref{rootdef}) we see that each root $E_{\rho}$ is uniquely associated with the vector $\rho$ of its charges under the Cartan subalgebra, and so one may identify $E_{\rho}$ with $\rho$.

In this notation, the roots of $\mathfrak{e}_7$ take the following form:
\begin{align}\label{roots}
&\left(\underline{\pm 1, \pm 1, 0,0,0,0},0\right)\\
&2\left(\pm 1, \pm 1, \pm 1, \pm 1, \pm 1, \pm 1, \pm \sqrt{2}\right) \label{root2}\\
&\left(0,0,0,0,0,0,\pm\sqrt{2}\right)
\end{align}
where in (\ref{root2}) we consider only charge vectors in which an even number of $+1$ appear.

In order to choose the vev of the Higgs field, we need to decompose $E_7 \to SU(5)\times SU(2)\times U(1)^2$. The dimension $133$ adjoint representation of $\mathfrak{e}_7$ decomposes as follows:
\bea
\mathfrak{e}_7 & \supset & \mathfrak{su}_{5}^{\rm GUT} \oplus \mathfrak{su}_{2}  \oplus \mathfrak{u}_{1} \oplus \mathfrak{u}_{1}\\ \nonumber
\textbf{133} & \rightarrow & (\textbf{24},\textbf{1})_{0,0} \oplus (\textbf{1},\textbf{3})_{0,0}  \oplus  2(\textbf{1},\textbf{1})_{0,0} \oplus ( (\textbf{1},\textbf{2})_{-2,1} \oplus  c.c.)  \\ \nonumber 
& & \oplus\ (\textbf{10},{\textbf{2}})_{1,0} \oplus (\textbf{10},{\textbf{1}})_{-1,1} \oplus ({\textbf{5}},{\textbf{2}})_{0,-1} \oplus ({\textbf{5}},{\textbf{1}})_{-2,0} \oplus ({\textbf{5}},{\textbf{1}})_{1,1}\oplus  c.c.
\eea
Let us look for a subset of $\mathfrak{su}(2)$ roots among the roots of $\mathfrak{e_7}$.
Notice that among the roots given in (\ref{roots}) we can identify two of them which add to zero. 
The generators associated to these roots will be the raising and lowering operators for the $\mathfrak{su(2)}$ subalgebra. 
We choose
\begin{subequations}
\begin{align}
E^{+}&:=E_{\frac{1}{2}(1,1,1,1,1,1,\sqrt{2})} \\
E^{-}&:=E_{-\frac{1}{2}(1,1,1,1,1,1,\sqrt{2})}
\end{align}
\end{subequations}
From the commutation rules for the root operators we have
\begin{align}
[E_{\alpha}, E_{\beta}]=\alpha\cdot \vec{H}
\end{align}
in the case that $\alpha+\beta=0$.
In the case of $E^\pm$ their commutator is given by
\begin{align}
P&:=[E^{+}, E^{-}]=\dfrac{1}{2}(H_1+H_2+H_3+H_4+H_5+H_6+\sqrt{2}H_7)
\end{align}
so that $\{E^+,E^-,P\}$ generates a $\mathfrak{su(2)}$ subalgebra of $\mathfrak{e_7}$.
\begin{subequations}
\begin{align}
[E^+, E^+]&=0\\
[E^-, E^-]&=0\\
[E^+, E^-]&=P
\end{align}
\end{subequations}
In the main text we use two particular linear combination of Cartan generators $Q_1$ and $Q_2$, that generate  the two Abelian factors in $\mathfrak{su}_{5}^{\rm GUT} \oplus \mathfrak{su}_{2}  \oplus \mathfrak{u}_{1} \oplus \mathfrak{u}_{1}$. These are
\begin{align*}
&Q_1=-\dfrac{1}{2}\left(H_1+H_2+H_3+H_4+H_5-H_6-2\sqrt{2} H_7\right)\\
&Q_2=-\dfrac{1}{2}\left(2H_6-\sqrt{2}H_7\right)
\end{align*}
With this assignment for the roots of the $SU(2)$ subgroup and the generators of the two $U(1)$s, we can also identify how all the other roots of $E_7$ split into representations of $SU(5)\times SU(2) \times U(1)\times U(1)$. We find

\begin{table}[h]
\begin{center}
\begin{tabular}{|c|c|c|}
\hline 
$E_7$ generator & $SU(5)\times SU(2)$ &$ Q_1, Q_2$ charges\\ 
\hline 
$(\underline{+1,-1,0,0,0},0,0)\oplus H_1, H_2, H_3, H_4$ & $(\mathbf{24},\mathbf{1})$ & (0,0) \\ 
$Q_1, Q_2$ cartans & $2(\mathbf{1},\mathbf{1})$ & (0,0) \\ 
$\rho_+,\rho_-, P$ & $(\mathbf{1},\mathbf{3})$ & (0,0) \\ 
$(0,0,0,0,0,0,\sqrt{2})$ and $\frac{1}{2}(-1,-1,-1,-1,-1,-1,\sqrt{2})$ & $(\mathbf{1},\mathbf{2})$ & (-2,1) \\ 
$\frac{1}{2}(1,1,1,1,1,1,-\sqrt{2})$ and $(0,0,0,0,0,0, -\sqrt{2})$ & $(\mathbf{1},\mathbf{2})$ & (2,-1) \\ 
$(\underline{1,0,0,0,0},1,0)$ and $\frac{1}{2}(\underline{1,-1,-1,-1,-1},1,-\sqrt{2})$ & $(\mathbf{5},\mathbf{2})$ & (0,-1) \\ 
$\frac{1}{2}(\underline{1,-1,-1,-1,-1},1,\sqrt{2})$ & $(\mathbf{5},\mathbf{1})$ & (-2,0) \\ 
$(\underline{1,0,0,0,0},-1,0)$ & $(\mathbf{5},\mathbf{1})$ & (1,1) \\ 
$(\underline{1,1,0,0,0},0,0)$ and $\frac{1}{2}(\underline{1,1,-1,-1,-1},-1,-\sqrt{2})$ & $(\mathbf{10},\mathbf{2})$ & (1,0) \\ 
$\frac{1}{2}(\underline{-1,-1,-1,1,1},-1,\sqrt{2})$ & $(\mathbf{10},\mathbf{1})$ & (-1,1) \\ 
\hline
\end{tabular}
\caption{Roots of $E_7$ and their charges under the subgroup $SU(5) \times SU(2) \times U(1)^2$.}
\end{center} 
\end{table}

\section{Local chirality and doublet-triplet splitting}
\label{ap:chiral}

In this appendix we provide further details on the computations regarding the local chirality for the models A and B, as follows from the discussion of section \ref{sec:flux}.

\subsection{Model A}

Using the explicit form of $q_R$ and $q_S$ that may be found in table \ref{t:sectors} we can write explicitly the chirality conditions \eqref{eq:chirals} for all the various sectors appearing in the model. These are
\begin{subequations}\label{eq:chirA}
\begin{align}
& q_Y \tilde N_Y - M_1 >0 \,, \qquad   q_Y = -\frac{1}{6}\,, \frac{2}{3}\,, -1 \label{eq:chirA1}\\&
q_Y \tilde N_Y -M_2 >0 \,, \qquad q_Y = \frac{1}{2}\,, -\frac{1}{3}\label{eq:chirA2}\\&
\left(|a|^2 -1\right) \left(2
   M_1+\frac{\tilde N_Y}{3}\right)+2\, \text{Re}[a] 
   \left(\frac{N_Y}{3}-2 N_1\right)=0\label{eq:chirA3}\\&
(|a|^2-1)\left(2
  M_1-\frac{\tilde N_Y}{2}\right)-2 \,\text{Re}[a] \left(2
   N_1+\frac{N_Y}{2}\right)>0\label{eq:chirA4}\\&
- \left(N_1+N_2 +\frac{N_Y}{3}\right)\text{Re}[(\mu_1^2+\mu_2^2)(a \mu_1^2 +b \mu_2^2)]+\frac{1}{2}\left (M_1+M_2 -\frac{\tilde N_Y}{3}\right) \hat \mu^4=0\label{eq:chirA5}\\&
- \left(N_1+N_2 -\frac{N_Y}{2}\right)\text{Re}[(\mu_1^2+\mu_2^2)(a \mu_1^2 +b \mu_2^2) ]+\frac{1}{2}\left (M_1+M_2 +\frac{\tilde N_Y}{2}\right) \hat \mu^4>0\label{eq:chirA6}
\end{align}\end{subequations}
where we defined $\hat \mu^4 = |a\,\mu_1^2 + b \,\mu_2^2|^2 -|\mu_1^2+\mu_2^2|^2$. 
From \eqref{eq:chirA1} and \eqref{eq:chirA2} we find two possible branches according to the sign of $\tilde N_Y$
\be
\tilde N_Y \leq 0 \quad \rightarrow \quad \left\{\begin{array}{l}M_1 < \frac{2}{3} \tilde N_Y\\[2mm] M_2 < \frac{1}{2} \tilde N_Y\end{array}\right.\,,
\quad \tilde N_Y > 0 \quad \rightarrow \quad \left\{\begin{array}{l}M_1 <- \tilde N_Y\\[2mm] M_2 <- \frac{1}{3} \tilde N_Y\end{array}\right.\,.
\ee
For each of these branches the whole system \eqref{eq:chirA} has solution and therefore it is possible to obtain (at least locally) the correct
chiral spectrum of the MSSM. This is in contrast to what happened in \cite{mrz15} where a solution was not possible. We can easily understand 
why this occurs by merely taking $a=b=1$ which was the particular case considered in \cite{mrz15}. If we impose \eqref{eq:chirA3} and 
\eqref{eq:chirA5} with $a=b=1$ then \eqref{eq:chirA4} and 
\eqref{eq:chirA6} reduce to $-N_1 >0$ and $N_1>0$ respectively and therefore the system does not allow for solutions.

\subsection{Model B}

For the model B we find a similar set of equations that are necessary to obtain the correct 4d chiral spectrum
\begin{subequations}\label{eq:chirB}
\begin{align}
& q_Y \tilde N_Y - M_1 >0 \,, \qquad   q_Y = -\frac{1}{6}\,, \frac{2}{3}\,, -1 \label{eq:chirB1}\\&
\frac{1}{2} \tilde N_Y -M_2 >0 \,, \label{eq:chirB2}\\&
\frac{1}{3} \tilde N_Y +M_2 =0 \,, \label{eq:chirB3}\\&
(|a|^2 -1) \left(2
   M_1+\frac{\tilde N_Y}{3}\right)+2 \text{Re}[a] 
   \left(\frac{N_Y}{3}-2 N_1\right)=0\label{eq:chirB4}\\&
(|a|^2-1)\left(2
  M_1-\frac{\tilde N_Y}{2}\right)-2 \text{Re}[a] \left(2
   N_1+\frac{N_Y}{2}\right)>0\label{eq:chirB5}\\&
- \left(N_1+N_2 +\frac{N_Y}{3}\right)\text{Re}[(\mu_1^2+\mu_2^2)(a \mu_1^2 +b \mu_2^2)] +\frac{1}{2}\left (M_1+M_2 -\frac{\tilde N_Y}{3}\right) \hat \mu^4>0\label{eq:chirB6}\\&
- \left(N_1+N_2 -\frac{N_Y}{2}\right)\text{Re}[(\mu_1^2+\mu_2^2)(a \mu_1^2 +b \mu_2^2) ]+\frac{1}{2}\left (M_1+M_2 +\frac{\tilde N_Y}{2}\right) \hat \mu^4>0\label{eq:chirB7}
\end{align}\end{subequations}

In this case the sign of $\tilde N_Y$ is fixed and there is only one branch of solutions to \eqref{eq:chirB1}-\eqref{eq:chirB3}, namely 
\be
\left\{\begin{array}{l}M_1 < -\tilde N_Y\\\tilde N_Y >0 \\ M_2 = -\frac{\tilde N_Y}{3}\end{array}\right.\,.
\label{condBmodel}
\ee
Here the whole system \eqref{eq:chirB} admits solutions even at the point $a=b=1$ 
and therefore it is also possible to obtain the correct chiral spectrum in 4d.

\section{Zero mode wavefunctions}
\label{ap:holoyuk}

In this appendix we present details of the computation of the zero modes wavefunctions in holomorphic and real gauge. We start by collecting
some data necessary for the computation of the Yukawa couplings in holomorphic gauge and then discuss how to obtain a solution of the full system of 
F-term and D-term equations in real gauge for all sectors appearing in the model.

\subsection{Wavefunctions in holomorphic gauge and Yukawa couplings}

Once the non-perturbative corrections are taken into account the Yukawa matrix has the following general form
\be\label{ap:yukres}
Y =  m_*^4 \pi^2 f_{abc}\,\text{Res}_p \left[\eta^a \eta^b h_{xy}\right],
\ee
where (at first order in $\eps$) $\eta$ is given by
\be
\eta = - i \Phi^{-1}\left[h_{xy}+i \epsilon\p_x \theta_0 \p_y \left (\Phi^{-1} h_{xy}\right)-i \epsilon\p_y \theta_0 \p_x \left (\Phi^{-1} h_{xy}\right)\right].
\ee
The explicit form of $\eta$ is different for each sector because of the different form of $\Phi$ and $h_{xy}$. Here we give the explicit expression for $h$ and $\eta$ for the various sectors, taking $\th_0$ as
\be
\th_0 = i (x \th_x +y \th_y)\,.
\ee
The result for each sector is

\subsection*{$\mathbf{(10,2)}_{1,0}$}
{\footnotesize
\bea
h^i_{\mathbf{10}} /\gamma_{\mathbf{10}}& = &   m_*^{3-i} (ax-y)^{3-i} \,,\\[2mm] \label{soleta10}
i \eta_{\mathbf{10}}^i/\g_{\mathbf{10}}^i & = &\frac{m_*^{3-i}}{\text{det}\Phi_{\mathbf{10}}}\left(
\begin{array}{c}
 -m  (a x-y)^{3-i} \\
 \mu _1^2  (a x-y)^{4-i} \\
\end{array}
\right)\\[2mm]\nonumber
&+&\eps\,\frac{ m_*^{3-i}}{(\text{det} \Phi_{\mathbf{10}})^3}
\left(
\begin{array}{c}
 -m  (a x-y)^{3-i} \\
 \mu _1^2  (a x-y)^{4-i} \\
\end{array}
\right)\left[m^3 \theta _y-2 \mu _1^4 (a x-y) \left(a \theta _y+\theta _x\right)\right]\\[2mm]\nonumber
&+&\eps\, \frac{m_*^{3-i}\left(a \theta _y+\theta
   _x\right) }{(\text{det} \Phi_{\mathbf{10}})^2} \left(
\begin{array}{c}
 -i (2 i-7) \mu _1^2\, m (a x-y)^{3-i} \\
 i  (a x-y)^{2-i} 
   \left((i-4) \mu _1^4 (y-a x)^2+(i-3) m^3 x\right) \\
\end{array}
\right)\,.
 \eea}

\subsection*{$\mathbf{(5,1)}_{-2,0}$}
\bea
h_{\mathbf{5,1}}/\g_{\mathbf{5,1}} & = & 1\\
i \eta_{\mathbf{5,1}}/\g_{\mathbf{5,1}} & = &  -\frac{1}{ \Phi_{\mathbf{5}_U}} + \eps\, \frac{2\mu^2(\th_{x} + a\, \th_{y})}{\Phi_{\mathbf{5}_U}^3} \,.
\eea

\subsection*{$\mathbf{(\bar 5,1)}_{-1,-1}$}

\bea
h^i_{\mathbf{\bar{5},1}}&=& m_*^{3-i}\left(\left(\mu _1^2+\mu
   _2^2\right) x+ \left(a \mu _1^2+b \mu _2^2\right)y\right){}^{3-i}\\[2mm]
i   \eta^{i}_{\mathbf{\bar{5},1}}/  \gamma^{i}_{\mathbf{\bar{5},1}}&=&-\frac{1}{\Phi_{\mathbf{\bar 5,1}}} h^i_{\mathbf{\bar{5},1}}+\eps\, \frac{h^i_{\mathbf{\bar 5,1}}}{\Phi_{\mathbf{\bar 5,1}}^3}(\theta _y \left(a \mu _1^2+b \mu _2^2\right)+\left(\mu _1^2+\mu _2^2\right)
   \theta _x)
\\[2mm]\nonumber
&+&\eps\,
\frac{m_* h^{i-1}_{\mathbf{\bar{5},1}}}{\Phi_{\mathbf{\bar 5,1}}^2}(3-i) 
   \left[\theta _x \left(a \mu _1^2+b \mu _2^2\right)-\left(\mu _1^2+\mu
   _2^2\right) \theta _y\right]
\eea

\subsection*{$\mathbf{(\bar 5,2)}_{0,1}$}

{\footnotesize
\bea
h^i_{\mathbf{\bar 5,2}} &= & m_*^{3-i} (a (x - x_0) - (y - y_0))^{3 - i}\,,\\[2mm]
i \eta^i_{\mathbf{\bar 5,2}}/\gamma^i_{\mathbf{\bar 5,2}}&=&\frac{m_*^{3-i}}{\text{det} \Phi_{\mathbf{\bar 5,2}}}\left(
\begin{array}{c}
 -m  \left(a \left(x-x_0\right)-y+y_0\right){}^{3-i} \\
 \left(a \left(x-x_0\right)-y+y_0\right){}^{3-i} \left(\mu
   _2^2 (b x-y)+\kappa \right) \\
\end{array}
\right)\\[2mm]\nonumber
&+&\eps\, \th_y \frac{m_* h^{i-1}_{\mathbf{\bar 5,2}}}{(\text{det} \Phi_{\mathbf{\bar 5,2}})^2}
\left(
\begin{array}{c}
 \mu _2^2 m \left[a b \left((2 i-7) x+x_0\right)-2 a (i-3) y+b
   \left(y-y_0\right)\right] \\
- \mu _2^4 (b x-y) \left[a b \left((i-4) x+x_0\right)-a (i-3) y+b
   \left(y-y_0\right)\right] \\
\end{array}
\right)\\[2mm]\nonumber
&+&\eps\, \th_y \frac{m_* h^{i-1}_{\mathbf{\bar 5,2}}}{(\text{det} \Phi_{\mathbf{\bar 5,2}})^2}
\left(
\begin{array}{c}
 2 a (i-3) \kappa  m \\
 \kappa \mu _2^2   \left[a b \left((7-2i) x-x_0\right)
  +b \left(y-y_0\right)\right] -a (i-3) (m^3 x-2 \kappa\mu_2^2 y -\kappa^2)\\
\end{array}
\right)\\[2mm]\nonumber
&+&\eps\, \th_x \frac{m_* h^{i-1}_{\mathbf{\bar 5,2}}}{(\text{det} \Phi_{\mathbf{\bar 5,2}})^2}
\left(
\begin{array}{c}
- \mu _2^2 m \left(a \left(x-x_0\right)-2 b (i-3) x+2 i y-7 y+y_0\right) \\
 \mu _2^4 (b x-y) \left(a \left(x-x_0\right)-b (i-3) x+(i-4)
   y+y_0\right)-(i-3) m^3 x \\
\end{array}
\right)
\\[2mm]\nonumber
&+&\eps\, \th_x \frac{ \kappa m_* h^{i-1}_{\mathbf{\bar 5,2}}}{(\text{det} \Phi_{\mathbf{\bar 5,2}})^2}
\left(
\begin{array}{c}
 2 (i-3)   m \\
   \mu _2^2 \left(a x-a x_0-2 b i x+6 b x+2 i y-7
   y+y_0\right)-(i-3) \kappa  \\
\end{array}
\right)\\[2mm]\nonumber
&+& \eps \,\th_y \frac{h^{i}_{\mathbf{\bar 5,2}}}{(\text{det} \Phi_{\mathbf{\bar 5,2}})^3} \left(
\begin{array}{c}
 -2 m \left(\mu _2^2 (b x-y)+\kappa \right) \left(-2 b \kappa  \mu _2^2+2 b
   \mu _2^4 (y-b x)+m^3\right) \\
 \left(-2 b \kappa  \mu _2^2+2 b \mu _2^4 (y-b x)+m^3\right) \left(2 \kappa 
   \mu _2^2 (b x-y)+\mu _2^4 (y-b x)^2+\kappa ^2+m^3 x\right) \\
\end{array}
\right)\\[2mm]\nonumber
&+& \eps \,\th_x \frac{h^{i}_{\mathbf{\bar 5,2}}}{(\text{det} \Phi_{\mathbf{\bar 5,2}})^3} \left(
\begin{array}{c}
 4 \mu _2^2 m \left(\mu _2^2 (b x-y)+\kappa \right){}^2 \\
 -2 \mu _2^2 \left(\mu _2^2 (b x-y)+\kappa \right) \left(2 \kappa  \mu _2^2 (b
   x-y)+\mu _2^4 (y-b x)^2+\kappa ^2+m^3 x\right) \\
\end{array}
\right)\,.
\eea}

Note that in the model A the sector $\mathbf{(\bar 5,2)}_{0,1}$ contains three families, so we must take $i=1,2,3$ in the expression above, while the sector $\mathbf{(\bar 5,1)}_{-1,-1}$ contains just one, so there we take $i=3$. In the model B the opposite happens, and so $i=1,2,3$ for the sector $\mathbf{(\bar 5,1)}_{-1,-1}$ and $i=3$ for the $\mathbf{(\bar 5,2)}_{0,1}$ sector.

\subsection{Wavefunctions in real gauge}

When computing the zero mode wavefunctions in real gauge we find that there is a great difference in the computation according to if the sector we are 
considering is charged or not under the T-brane background. Because of this we shall separate the discussion starting with sectors not affected by the 
T-brane background. 
\subsubsection*{Sectors not affected by the T-brane background}

In these sectors which do not feel the effect of the non-commutativity of the background Higgs field it is possible to solve exactly for the wavefunctions using the techniques already employed in \cite{afim11,fimr12}. The F-term and D-term equations may be compactly rewritten as a Dirac-like equation 
\be
\left(
\begin{array}{cccc}
0 & D_x & D_y & D_z \\
-D_x & 0 & -D_{\bar{z}} & D_{\bar{y}} \\
-D_y & D_{\bar{z}} & 0 & -D_{\bar{x}} \\
-D_z & -D_{\bar{y}} & D_{\bar{x}} & 0
\end{array}
\right)
\left(
\begin{array}{c}
0 \\ \\  \vec{\vphi}_{U} \\ \quad
\end{array}
\right)\, =\, 0  
\label{Dirac5}
\ee
where we defined the covariant derivatives
\be
D_{x} \, =\, \p_{x} + \oh(q_R \bar{x} - q_S \bar{y}) \qquad D_{y} \, =\, \p_{y} - \oh (q_R \bar{y}  + q_S \bar{x}) \qquad D_{z}\, =\, 2i (\tilde \mu_a^2 \bar{x}-\tilde \mu_b^2 \bar{y}) 
\label{covar}
\ee
and $D_{\bar m}$ are their conjugate. In writing the covariant derivatives we took the following gauge connection
\be\label{gaugetot}
A=\frac{i}{2}Q_R(yd\bar y-\bar ydy-xd\bar x+\bar xdx)+\frac{i}{2}Q_S(xd\bar y-\bar ydx+yd\bar x-\bar xdy)-\frac{i}{2}m^2c^2P(xd\bar x-\bar xdx)\,,
\ee
which gives the flux 
\be
F=iQ_R(dy\wedge d\bar y-dx\wedge d\bar x)+iQ_S(dx\wedge d\bar y+dy\wedge d\bar x)+im^2c^2P dx\wedge d\bar x\,.
\ee
The effect of the fluxes in every sector is different and reflected in the values of the constants $q_R$ and $q_S$ which appear in the covariant derivatives. The various values of $q_R$ and $q_S$ of the different MSSM fields are listed in table \ref{t:sectors}. We can follow the strategy outlined in \cite{fimr12} to find a solution for the previous system of differential equations and the result is
\be
\vec \varphi = \left(\begin{array}{c}-\frac{i \zeta}{2 \tilde \mu_a} \\ \frac{i (\zeta-\lambda)}{2 \tilde \mu_b}\\1\end{array}\right) \chi (x,y) 
\ee
where
\be
\chi (x,y) = e^{\frac{q_R}{2} (x \bar x -y \bar y)-q_S\text{Re} (x \bar y)+ (\mu_a x+ \mu_b y)(\zeta_1 \bar x - \zeta_2 \bar y)}\,f(\zeta_2x+\zeta_1y)
\ee
we have defined 
\be
\zeta = \frac{\tilde \mu _a \left(4 \tilde \mu _a\tilde  \mu _b+\lambda  q_S\right)}{\tilde \mu _a q_S+\tilde \mu _b \left(\lambda +q_R\right)}\,,\quad 
\zeta_1 = \frac{\zeta}{\tilde \mu_a} \,, \quad \zeta_2 = \frac{\zeta-\lambda}{\tilde \mu_b}\,,
\ee
and $\lambda$ is defined as the lowest solution of the cubic equation
\be\label{eq:cub1}
-\lambda ^3+4 \lambda  \mu_a^2+4 \lambda  \mu_b^2+\lambda  q_R^2-4
\mu_a^2 q_R+4 \mu_b^2 q_R+\lambda  q_S^2+8 \mu_a
   \mu_b q_S = 0\,.
\ee
This general solution applies to any sector whose matter curve goes through the origin. The effect of a non-zero separation 
(which affects only the $\bar{\mathbf 5}_{-1,-1}$ sector) can be easily taken into account by performing a shift in the $(x,y)$ plane
\be
x \rightarrow x- x_0\,, \quad y\rightarrow y-y_0\,.
\ee
However by simply performing the shift in the scalar wavefunction $\chi$ we would obtain a solution for a shifted gauge field $A$.
This may be easily remedied by a suitable gauge transformation
\be
A (x-x_0 , y-y_0) = A(x,y) + d \psi\,,
\ee
with
\be\label{gaugepar}
\psi=\frac{i}{2}Q_R(y_0\bar y-\bar y_0y-x_0\bar x+ x_0x)+\frac{i}{2}Q_S(x_0\bar y-\bar y_0x+y_0\bar x-\bar x_0y)-\frac{i}{2}m^2c^2P(x_0\bar x-\bar x_0x).
\ee
Therefore the general shifted solution may be written as 
\be
 {\vec \varphi} = \left(\begin{array}{c}-\frac{i \zeta}{2 \tilde \mu_a} \\ \frac{i (\zeta-\lambda)}{2 \tilde \mu_b}\\1\end{array}\right) e^{-i \psi}
\chi (x-x_0,y-y_0) \,.
\ee

\subsubsection*{Sectors  affected by the T-brane background}

The presence of the the T-brane background greatly affects the sectors charged under it and in particular as we are now going to show it turns out prohibitive 
to find a simple solution to the zero modes equations of motion. However in particular region in the space of parameters, more precisely when the 
diagonal terms in the Higgs background are negligible compared to the off-diagonal ones, great simplifications occur in the zero-mode equations and 
a solution may be easily obtained. 

The sectors affected by the T-brane background are the $(\mathbf{10,2})_{1,0}$ and the $(\mathbf{\bar 5,2})_{0,1}$. Since the difference between the 
two appears in the diagonal entries of the Higgs background and we are going to assume that these contributions are negligible we shall discuss
them at the same time in the following.

The general form of the wavefunctions for the sectors charged under the T-brane is the following one

\be
\left(
\begin{array}{c}
a_{\bar{x}}\\ a_{\bar{y} }\\ \vphi_{xy}
\end{array}
\right)
\, =\, 
\vec{\vphi}_{{10}^+} E_1^+ + \vec{\vphi}_{{10}^-} E_1^-\,.
\ee

The zero-mode equations take the same form of \eqref{Dirac5} when written in terms of 
\be
a = \bmat{c} a^+ \\
a^-
\emat\,, \qquad  \qquad
\vphi = \bmat{c} \vphi^+ \\
\vphi^-
\emat\,.
\label{doubs}
\ee
Following \cite{fmrz13} we will start by looking for a general solution of the F-term equations and eventually impose the D-term equations on this
solution. While the first step may be done for a general choice of the parameters entering in the Higgs background the latter turns out to be feasible if we
restrict to the particular case in which the diagonal terms in the Higgs background are negligible as opposed to the off-diagonal ones.

For sake of notational simplicity we will consider the case in which the primitive fluxes are vanishing and reinstate them at the end of the computation.
Then the general solution to the F-terms is 
\begin{subequations}
\label{Fsol}
\begin{align}
\label{asol}
a & =  e^{fP/2} \bar\p \xi\\
\label{phisol}
\vphi & =  e^{fP/2}\left(h - i \Psi \xi \right) 
\end{align}
\end{subequations} 
where $\xi$ and $h$ are both doublets whose components we denote as $\xi^\pm$ and $h^\pm$ and $P$ and $\Psi$ when acting on doublets may
be represented as
\be
P = \bmat{cc} 1 & 0 \\
0 & -1
\emat \,,\qquad  \qquad
\Psi = \bmat{cc} \tilde \mu^2F(x,y) & m \\
m^2 x & \tilde \mu^2f(x,y)
\emat\,.
\ee
The explicit form of $\tilde \mu^2 F(x,y)$ is different in the two sectors that we are considering in this section but it will be unimportant in the upcoming
discussion as we will choose these terms to be negligible.

We may now solve \eqref{Fsol} for $\xi$ obtaining
\be
\xi = i \Psi^{-1} \left( e^{-fP/2}\vphi - h\right)\,,
\label{xisol}
\ee
and plug this solution in the D-term equations for the zero-modes which therefore become an equation in $\xi$ and $h$
\be
\p_x \p_{\bar x} \xi + \p_y \p_{\bar y} \xi + \p_x f P \p_{\bar x} \xi - i \Lambda^\dagger \left(h - i \Psi \xi \right)  = 0   \,.
\label{dterm2} 
\ee 
Note that in writing \eqref{dterm2}  we have used that the function $f$ does not depend on $(y, \bar y)$ and we have defined 
\be
\Lambda = e^{f P } \Psi e^{-fP} = \bmat{cc} \tilde \mu^2F(x,y) & m e^{2f} \\
m^2 x e^{-2f} & \tilde \mu^2F(x,y)
\emat\,.
\ee
While \eqref{dterm2} depends on both $\xi$ and $h$ it is possible to write it as an equation for one single doublet $U$ defined as
\be
U = e^{-fP/2} \vphi \,, \qquad \rightarrow \qquad \xi = i \Psi^{-1} (U-h)\,.
\ee
When written in terms of $U$ \eqref{dterm2} becomes
\be
\p_x \p_{\bar x} U + \p_y \p_{\bar y} U 
-  (\p_x \Psi) \Psi^{-1}  \p_{\bar x} U  + (\p_y \Psi) \Psi^{-1} \p_{\bar y} U
+  \p_x f \Psi P \Psi^{-1} \p_{\bar x} U -  \Psi\Lambda^\dagger U = 0\,.
\label{uterm}
\ee

We have managed therefore to translate the full set of zero-mode equations to a system of partial differential equations for the doublet $U$. In general 
this system is coupled and therefore finding a solution turns out to be very involved. However when taking $\tilde \mu^2 \ll m^2$ the system decouples
and may be easily solved. Since no localised solution of this system for $U^+$ exists we will set to zero henceforth. Then if we Taylor expand $f$ near the Yukawa
point as $f = \log c + c^2 m^2 x \bar x$ we find that $U^- = e^{\lambda x \bar x}$ where $\lam$ the lowest solution to $ c^2 \lam^3 + 4c^4 m^2 \lam^2 - m^4\lam=0$. Using this the solution to the zero mode equations is simply
\be
\vec{\vphi}_{^+}^j \, =\, \gamma^j
\left (\begin{array}{c}
\frac{i\lam}{m^2}\\
0 \\
0 \end{array}\right )
 e^{f/2} \chi^j \,,
 \qquad \quad 
\vec{\vphi}_{^-}^j \, =\, \gamma^j
\left (\begin{array}{c}
0 \\
0 \\
1 \end{array}\right )
e^{-f/2} \chi^j \,,
\ee
where $e^{f/2} = \sqrt{c}\, e^{m^2c^2x\bar{x}/2}$ and $\chi^j \, =\, e^{\lam x \bar x}\, g_j (y)$, with  $g_j$ holomorphic functions of $y$. 

It is easy to generalise the computation when extra primitive fluxes are present. Following a similar procedure we obtain a solution which now is
\be
\label{phys10} 
\vec{\vphi}_{^+}^j \, =\, \gamma^j
\left (\begin{array}{c}
\frac{i\lam}{m^2}\\
-\frac{i\lam\zeta}{m^2} \\
0 \end{array}\right )
 e^{f/2} \chi^j 
 \qquad \quad 
\vec{\vphi}_{-}^j \, =\, \gamma^j
\left (\begin{array}{c}
0 \\
0 \\
1 \end{array}\right )
e^{-f/2} \chi^j 
\ee
where $\lam$ is the lowest (negative) solution to 
\be\label{cub10}
m^4 (\lambda -q_R)+\lambda c^2 \left( c^2 m^2 (q_R-\lambda )-\lambda ^2+q_R^2+q_S^2\right)=0
\ee
and $\zeta = -q_S/(\lam-q_R)$. The scalar wavefunctions $\chi$ are
\be
\chi^j \, =\,e^{\frac{q_R}{2}(|x|^2-|y|^2)-q_S (x \bar y +y\bar x)+\lam x (\bar x - \zeta \bar y)} \, g_j (y +\zeta x)
\label{wave10p}
\ee
where $g_j$ holomorphic functions of $y +\zeta x$, and $j=1,2,3$ label the different zero mode families. The family functions we choose to adopt
are
\be
g_j\, =\, m_*^{3-j}(y +\zeta x)^{3-j}.
\label{holorep}
\ee

Note that in neglecting the diagonal terms in the Higgs background we may also discard the effect of the separation of the Yukawa points. If however we
consider the case $\kappa,\mu_2 \ll m$ with $\kappa/\mu_2^2 = \nu$ finite we find that the down Yukawa point is located at $(x_0,y_0) = (0,\nu/2)$.
We may follow the same strategy as in the previous section and obtain the solution by simply performing a shift and the result is

\be
\vec \varphi^i = \gamma^i \left(\begin{array}{c}\frac{i \lambda}{m^2}\\-i\frac{\lambda \zeta}{m^2}\\0\end{array}\right)e^{i\tilde\psi +f/2} \chi^i(x,y-\nu/2) E^++\gamma^i \left(\begin{array}{c}0\\0\\1\end{array}\right)e^{i\tilde\psi-f/2} \chi^i(x,y-\nu/2) E^-
\ee
where $\tilde \psi$ is 
\be\label{gaugepar2}
\tilde\psi=\frac{i}{2}Q_R(\nu\bar y/2-\bar \nu y/2)+\frac{i}{2}Q_S(\nu\bar x/2-\bar \nu x/2)\,,
\ee
and the definitions of $\chi$, $\zeta$ and $\lambda$ are unchanged.

\subsubsection*{Non-perturbative corrections - sectors not affected by T-branes}

The computation of the first order correction to the wavefunction is similar to the one already considered in \cite{fmrz13}. The zero-mode equations
are
\bea
\bar\partial_{\langle A\rangle}a&=&0\,,\\
\bar\partial_{\langle A\rangle}\varphi+i[\langle \Phi\rangle,a]+\eps\,\partial\th_0\,\w\,\p_{\langle A\rangle}a&=&0\,,\\
\omega\,\w\,\p_{\langle A\rangle}a-\frac{1}{2}[\langle \bar\Phi\rangle,\varphi]&=&0\,.
\eea
We find it possible to solve for the first order correction to the wavefunctions and the result is
\be\label{corr1}\begin{split}
\vec\vphi^{(1)}\, &=\, \gamma\left(\begin{array}{c}-\frac{i \zeta}{2 \tilde \mu_a} \\ \frac{i (\zeta-\lambda)}{2 \tilde \mu_b}\\1\end{array}\right) \, 
e^{\frac{q_R}{2} (x \bar x -y \bar y)-q_S\text{Re} (x \bar y)+ (\tilde\mu_a x+\tilde  \mu_b y)(\zeta_1 \bar x - \zeta_2 \bar y)}\, \Upsilon\,.
\end{split}\ee
The function $\Upsilon$ that controls the $\CO(\eps)$ correction is
\be\begin{split}
\Upsilon &= \frac{1}{4} (\zeta_1 \bar x -\zeta_2 \bar y )^2 (\theta_y \mu_a - \theta_x\mu_b) f (\zeta_2 x + \zeta_1 y)+ \frac{1}{2} (\zeta_1 \bar x -
\zeta_2 \bar y)  (\zeta_2 \th_y - \zeta_1 \th_x)f'(\zeta_2 x + \zeta_1 y)+\\
&+ \left[\frac{\delta_1}{2} (\zeta_1 x - \zeta_2 y)^2+\delta_2  (\zeta_1 x - \zeta_2 y) (\zeta_2 x + \zeta_1 y)\right]f(\zeta_2 x + \zeta_1 y)\,,
\end{split}\ee
where
\be\begin{split}
\delta_1 & = \frac{1}{(\zeta_1^2 + \zeta_2^2)^2} \left[\bar \th_x (q_S \zeta_1 -q_R \zeta_2)+ \bar \th_y (q_R \zeta_1 +q_S \zeta_2)\right]\,,\\
\delta_2 & = \frac{1}{(\zeta_1^2 + \zeta_2^2)^2} \left[\bar \th_x (q_R \zeta_1 +q_S \zeta_2)- \bar \th_y (q_S \zeta_1 -q_R \zeta_2)\right]\,.
\end{split}\ee

Similarly to the order zero in $\eps$ case it is possible to obtain the solution when there is a non-zero separation between the Yukawa points by
performing a shift in the coordinates and a suitable gauge transformation. Because of this similarity we refrain from displaying the
result explicitly.

\subsubsection*{Non-perturbative corrections - sectors  affected by the T-brane background}

As already mentioned in the main text the mere structure of the first order correction to the wavefunctions is sufficient to ensure that no corrections
at $\CO(\eps)$ are present in the kinetic terms. Here we shall demonstrate how this structure arises without explicitly computing the first order 
correction as this is unnecessary to compute the kinetic terms.

We start by consider the case when the primitive fluxes are absent. The solution to the F-term equations at first order in $\eps$ is
\begin{subequations}
\label{Fsolnp}
\begin{align}
\label{asolnp}
a & =  g\, \bar\p \xi\\
\label{phisolnp}
\vphi & =  g \left(h - i \Phi \xi - \eps \p \th_0  \wedge \p \xi \right) \, =\, g\, U\, dx \wedge dy
\end{align}
\end{subequations}
with
\be
g=\left ( \begin{array}{cc}
e^{f/2}&0\\
0&e^{-f/2}\end{array}\right )\,,
\ee
where $\Phi$ is different in the two sectors and may be found in \eqref{eq:PhiA}.
We expand the doublet $U$ in $\eps$
\be
U\, =\, U^{(0)} + \eps\, U^{(1)} + \, \CO(\eps^2)
\ee
where $U^{(0)}$ was computed previously
\be
 U^{(0)}_- \, =\,  e^{\lambda x\bar{x}} h (y) \qquad \qquad  U^{(0)}_+ \, =\, 0.
\ee
Then, one may solve for $\xi$ from (\ref{phisolnp}) as 
\be
\begin{array}{c}
\xi\, =\, \xi^{(0)} +  i \epsilon \Phi^{-1} \left[U^{(1)} + \partial_x\th_0 \p_y\xi^{(0)} - \p_y\th_0\p_x\xi^{(0)} \right] + \mathcal O(\epsilon^2) \\
 \xi^{(0)} \, =\, i \Phi^{-1} (U^{(0)} - h)
\end{array}
\label{solxinp}
\ee
and then solve for $U^{(1)}$ by plugging in this expression into the D-term for the fluctuations (\ref{treeD}). This yields $U^{(1)}_- =0$, in the limit 
$\tilde \mu^2\ll m^2$. Thus, we find the following structure
\be
\xi_+ \, =\, \xi_+^{(0)} + 0 + \CO(\eps^2) \qquad \quad \xi_- \, =\,  \eps\, \xi_-^{(1)} + \CO(\eps^2).
\ee
This is actually sufficient to prove that the solution when taking into account the first order correction in $\eps$ has the following form
\be
\vec{\vphi}_{+} \, =\, 
\left(\begin{array}{c} \bullet \\ \bullet \\ 0\end{array}\right)
 + \eps
\left(\begin{array}{c} 0 \\ 0 \\ \bullet\end{array}\right) +\CO(\eps^2)
 \qquad 
\vec{\vphi}_{-} \, =\,
\left(\begin{array}{c} 0 \\ 0 \\ \bullet\end{array}\right)
 + \eps
 \left(\begin{array}{c} \bullet \\ \bullet \\ 0\end{array}\right)+\CO(\eps^2)
\ee
This is sufficient for the argument outlined in the main text regarding the computation of the kinetic terms at order $\eps$. Moreover by following a similar
procedure it is possible to show that this continues to hold if primitive fluxes are taken into account and the shift form the origin is taken into account.
\subsection{Holomorphic Yukawa matrix}\label{app:YukA}

Let us give the explicit expressions for the down-type Yukawas  that arises from the residue formula (\ref{yukres}), for the case of the model A. Unlike in (\ref{hyukD}) the expression below will be given to all orders in the parameter $a-b$. We obtain that
\be
Y_{D/L} = \left(\begin{array}{ccc}0 & Y^{(12)} & Y^{(13)} \\ Y^{(21)} & Y^{(22)} & Y^{(23)}\\Y^{(31)} & Y^{(32)} &Y^{(33)}\end{array}\right)
+ \mathcal{O} (\eps^2)\,,
\ee
where
\be
Y^{(12)} =  \frac{2 \pi ^2 \tilde \kappa  \epsilon  \gamma _{5,2} \gamma
   _{10,1} \gamma_D \left(\theta _y (a+b d)+(d+1) \theta_x\right)}{(d+1)^4  \rho_m\rho _{\mu }^3}\,,
\ee
\be\begin{split}
Y^{(13)} &= -\frac{\pi ^2 \gamma _{5,3} \gamma _{10,1} \gamma _D}{(d+1)^5 \rho_m^{5/2} \rho _{\mu }^{5}}\left[\epsilon(d+1)^2 \rho_m^{3/2}   \rho _{\mu }^6 \left(\theta _y (a+b
   d)+(d+1) \theta _x\right)+2 (d+1)^2 \tilde \kappa ^2 \rho_m^{3/2} \rho _{\mu }^2\right.\\
&\left.   -2 d \tilde \kappa  \epsilon  (a-b) \rho _{\mu }^4 \left(\theta _y
   \left(a \left(2 d^2+7 d-1\right)+3 b (d-1)
   d\right)+\left(5 d^2+4 d-1\right) \theta _x\right)\right]\,,
\end{split}\ee
\be
Y^{(21)} = \frac{\pi ^2 \tilde \kappa  \epsilon  \gamma _{5,1} \gamma _{10,2}
   \gamma _D \left(\theta _y (a+b d)+(d+1) \theta
   _x\right)}{(d+1)^4 \rho_m \rho _{\mu }^3}\,,
\ee
\be\begin{split}
Y^{(22)} &= -\frac{\pi ^2 \epsilon  \gamma _{5,2} \gamma _{10,2} \gamma
   _D}{(d+1)^5 \rho_m^{5/2} \rho _{\mu }^2}\left[(d+1)^2 \rho_m^{3/2} \left(\theta _y (a+b d)+(d+1) \theta _x\right)\right.\\
   &-\left.d (a-b) \rho _{\mu } \left(\theta _y (-a (d-8)+10 b
   d+b)+9 (d+1) \theta _x\right)\right]\,,
\end{split}\ee
\be\begin{split}
Y^{(23)} &=\frac{\pi ^2 \gamma _{5,3} \gamma _{10,2} \gamma
   _D}{(d+1)^6 \rho_m^4 \rho _{\mu }^2}\left[\epsilon\,d^2 (d+1)^2 \rho_m^{3/2}\rho _{\mu }^2    (a-b) \left(\theta
   _y (a (d+4)+b (2 d-1))+3 (d+1) \theta _x\right)\right.\\
   &- \left.  6 \tilde \kappa d^2 \epsilon  (a-b)^2 \rho
   _{\mu }^3 \left(\theta _y (a (d (d+8)+2)+b d (6
   d+1))+(d+1) (7 d+2) \theta _x\right)\right.\\
   &+ \left.\tilde \kappa (d+1)^4 \rho_m^3 -6 d (d+1)^2 \tilde \kappa ^2 \rho_m^{3/2} (a-b) \rho _{\mu }\right]\,,
\end{split}\ee
\be\begin{split}
Y^{31} &= -\frac{\pi ^2 \epsilon  \gamma _{5,1} \gamma _{10,3} \gamma
   _D}{(d+1)^5 \rho_m^{5/2} \rho _{\mu }^2}\left[(d+1)^2 \rho_m^{3/2} \left(\theta _y (a+b d)+(d+1) \theta _x\right)\right.\\
   &-\left.2 \tilde \kappa  (a-b) \rho _{\mu } \left(\theta _y (a (2
   d-1)+b ((d-3) d-1))+(d-2) (d+1) \theta _x\right)\right]\,.
\end{split}\ee
\be\begin{split}
Y^{32} &= -\frac{\pi ^2  \epsilon \,d \gamma _{5,2} \gamma _{10,3}
    \gamma _D\,(a-b)}{(d+1)^6 \rho_m^4}\left[(d+1)^2 \rho_m^{3/2} \left(\theta _y (-a (d-2)+4 b d+b)+3 (d+1)
   \theta _x\right)\right.\\
   &\left.-2 \tilde \kappa  (a-b) \rho _{\mu } \left(\theta _y \left(d^2
   (19 b-6 a)+7 d (a+b)+a\right)+(d+1) (13 d+1) \theta
   _x\right)\right]
\end{split}\ee
\be\begin{split}
Y^{(33)}&=-\frac{\pi ^2 \gamma _{5,3} \gamma _{10,3} \gamma
   _D}{(d+1)^7 \rho_m^{11/2} \rho _{\mu }}\left[(d+1)^6\rho_ m^{9/2}-2 d (d+1)^2 \tilde\kappa \rho_ m^{3/2} (a-b) \rho _{\mu }
   \left(6 d \tilde \kappa  (b-a) \rho _{\mu }+(d+1)^2
   \rho_m^{3/2}\right) \right.\\&
   -2\epsilon \,d^2 (d+1)^2 \rho_m^{3/2}   (a-b)^2 \rho _{\mu }^3
   \left(\theta _y (a (1-(d-3) d)+b d (5 d+2))+(d+1) (4
   d+1) \theta _x\right)\\& 
   \left.+4 d^3 \tilde \kappa  \epsilon  (a-b)^3 \rho _{\mu }^4 \left(\theta
   _y (a ((19-7 d) d+11)+3 b d (11 d+6))+(d+1) (26 d+11)
   \theta _x\right)\right]\,.
\end{split}\ee

\section{Details of the model B}
\label{ap:modelB}

In this appendix we collect some details regarding the model B that we omitted in the main text. We start by writing down the full form of the Yukawa
matrices in the case of zero separation between the Yukawa points and then write the formulas for the various normalisation factors entering in the model.

\subsection{Yukawa matrices}
The only difference between the model A and the model B will appear in the Yukawa matrices for the down quarks and leptons, since these are the sectors that
involve a different matter assignment. In the case of the B model such matrices have the form
\be
Y_{D/L} = \left(\begin{array}{ccc}0 & 0 & Y^{(13)}\\0& Y^{(22)} & Y^{(23)} \\ Y^{(31)} & Y^{(32)} &Y^{(33)}\end{array}\right)+ \mathcal O(\eps^2)\,,
\ee
where
\be
Y^{(13)}=-\g_{10,1}\g_{5,3}\g_D\frac{\pi ^2 \epsilon  \left(a \theta _y+b d \theta _y+d \theta
   _x+\theta _x\right)}{(d+1)^3 \rho _{\mu }^2 \rho _m}\,,
   \ee
\be
Y^{(22)} = \g_{10,2}\g_{5,2}\g_D\frac{\pi ^2 \epsilon  \,\theta _y \sqrt{a^2+2 a b d+d \left(b^2
   d+d+2\right)+1}}{(d+1)^3 \rho _{\mu }^2 \rho _m}\,,
\ee
\bea
Y^{(23)} &=& \frac{\pi ^2 d \epsilon\,  (a-b)\g_{10,2}\g_{5,3}\g_D }{(d+1)^4 \rho _m^{5/2}}\left[\theta
   _x\left(3 d^2+5 d+2\right) \right.\\\nonumber&+&\left.\theta _y \left(a \left(d^2+4
   d+2\right)+b d (2 d+1)\right)\right]
\eea
\bea
Y^{(31)}&=&\frac{\pi ^2 \epsilon  (a+b d)\g_{10,3}\g_{5,1}\g_D }{(d+1)^3 \rho _{\mu }^2 \rho _m
   \left(a^2+2 a b d+\left(b^2+1\right) d^2+2 d+1\right)}\left[\theta _x(d+1)  (a+b d)\right.\\[2mm]\nonumber
   &-&\left.\theta _y \left(a^2+2 a b d+\left(b^2+2\right) d^2+4 d+2\right)\right]\,,
\eea
\bea
Y^{(32)}& =& \frac{\pi ^2 d \epsilon  \sqrt{a^2+2 a b d+\left(b^2+1\right) d^2+2
   d+1} \g_{10,3}\g_{5,2}\g_D }{(d+1)^4 \rho _m^{5/2}}\\\nonumber&\times & \left[\theta _y (-a d+2 b d+b)+(d+1) \theta
   _x\right]\,,
\eea
\be
Y^{(33)} =-\frac{\pi ^2 \g_{10,3}\g_{5,3}\g_D}{(d+1) \rho _{\mu } \rho _m}\left[ 1+ \frac{2 \pi ^2 d^2 \epsilon  (a-b)^2 \rho _{\mu }^3 \left(\theta _y
   (a (d+2)+b d)+2 (d+1) \theta _x\right)}{(d+1)^3 \rho _m^3}\right]\,.
\ee

\subsection{Normalisation factors}
Following the prescription described in the main text it is possible to compute the normalisation factors for the model B and the result is
\begin{subequations}\label{normB}
\begin{align}
|\gamma_{U}|^2=&-\frac{4}{ \pi^2 m_*^4}\frac{ \left(-2 \zeta _1 \tilde{\mu }_a-q_R\right) \left(-2 \zeta _2 \tilde{\mu }_b-q_R\right)+\left(\zeta _2 \tilde{\mu }_a-\zeta
   _1 \tilde{\mu }_b+q_S\right){}^2}{\zeta_1^2+\zeta_2^2+4 } \\
   |\gamma_{D}|^2=&\,-\frac{c}{m_*^2\pi^2(3-j)!}\frac{1}{\frac{1}{2\text{Re}[\lam_{5}]+q_R(1+|\zeta_{5}|^2)-|m|^2 c^2}+\frac{c^2|\lam_{5}|^2}{|m|^4}\frac{1}{2\text{Re}[\lam_{5}]+q_R(1+|\zeta_{5}|^2)+|m|^2 c^2}}\\
|\gamma_{10, j}|^2=&\,-\frac{c}{m_*^2\pi^2(3-j)!}\frac{1}{\frac{1}{2\text{Re}[\lam_{10}]+q_R(1+|\zeta_{10}|^2)-|m|^2 c^2}+\frac{c^2|\lam_{10}|^2}{|m|^4}\frac{1}{2\text{Re}[\lam_{10}]+q_R(1+|\zeta_{10}|^2)+|m|^2 c^2}}\left (\frac{q_R}{m_*^2}\right )^{4-j}\\
|\gamma_{5,j}|^2=&-\frac{2^{j-1}\, \mathcal N^{3-j}}{ \pi^2 (3-j)!m_*^{2(5-j)}}\frac{\left(2 \text{Re}[\zeta _1 \tilde{\mu }_a]+q_R\right) \left(2 \text{Re}[\zeta _2 \tilde{\mu }_b]+q_R\right)+\left|\zeta _2 \tilde{\mu }_a-\zeta^*
   _1 \tilde{\mu }^*_b+q_S\right|^2}{\zeta_1^2+\zeta_2^2+4 }  \,,
\end{align}
\end{subequations}
where
\be
\mathcal N = \frac{\left(2 \text{Re}[\zeta _1 \tilde{\mu }_a]+q_R\right) \left(2 \text{Re}[\zeta _2 \tilde{\mu }_b]+q_R\right)+\left|\zeta _2 \tilde{\mu }_a-\zeta^*
   _1 \tilde{\mu }^*_b+q_S\right|^2}{\left[2\left(|\zeta _1|^2+|\zeta _2|^2\right) \text{Re}\left[\zeta _1 \mu _a-\zeta _2 \mu _b\right]+2\text{Re}[\zeta ^*_1 \zeta _2]\,
   q_S-\left(|\zeta _2|^2-|\zeta _1|^2\right) q_R\right]}\,.
\ee


\end{document}